%% file: main.tex
\documentclass[%
 preprint,
superscriptaddress,
 amsmath,amssymb,
 aps, prx,
]{revtex4-2}
\input{revtex_preamble}

\begin{document}

\title{\textbf{Objective Nonadiabatic Quantum Dynamics}}
\author{Shivam Sharma}
\email{shivams@illinois.edu}
\affiliation{Materials Research  Laboratory, University of Illinois, Urbana-Champaign, Urbana, IL, 61801, USA}
\affiliation{Mechanical Science and Engineering, University of Illinois, Urbana-Champaign, Urbana, IL, 61801,USA}
\author{Richard D. James}
\email{james@umn.edu}
\affiliation{ Department of Aerospace Engineering and Mechanics,\ University of Minnesota, Minneapolis, MN, 55455, USA}%

\begin{abstract}
Far-from-equilibrium behavior in driven matter emerges from the coupled evolution of electronic excitations, nuclear motion and macroscopic deformation. Describing this interplay across quantum, atomistic and macroscopic scales in spatially extended materials remains challenging. The conventional symmetry reductions are based on static transformations and are generally not dynamically compatible with a time-dependent macroscopic motion. In this work, we introduce objective non-adiabatic quantum dynamics (OQD), an exact symmetry reduction of non-adiabatic quantum dynamics (NAQD) for systems undergoing prescribed macroscopic motion. Only the finite number of nuclei and electronic degrees of freedom are explicitly propagated, while all other nuclei and electrons filling all of space satisfy exactly the equations of NAQD. The construction employs discrete groups of Euclidean isometries, consisting of orthogonal transformations and translations with affine time-dependence, and exploits frame indifference together with permutation invariance of identical nuclei. We derive the accompanying group transformation of the many-body electronic wavefunction and prove that an initially group-equivariant electronic state remains equivariant under time evolution. We further show that this invariant structure is inherited by time-dependent Hartree–Fock and by time-dependent full configuration interaction within a chosen orbital space. For simulation, we adopt a Lagrangian approach, where the deformation enters through a metric kinetic operator and transformed interaction potentials. We implement the resulting equations using in-house real-space finite difference discretization of real-time time-dependent density functional theory (RT-TDDFT). As an application of the method, we simulate rapid compression of warm dense hydrogen from $0.6$ to $1.0$ g/cm$^3$ at two rates. At slower rate the electrons remain close to the instantaneous Born-Oppenhiemer (BO) ground state, whereas faster rate leaves five times as many electrons outside BO occupied manifold with hotter nuclei and more dissociated molecules. In both case we observe an insulator-to-metal transition.   



\end{abstract}

\maketitle


\section{\label{sec:intro}Introduction}

The coupled and non-equilibrium dynamics of electrons and nuclei in driven material systems are ubiquitous in condensed matter systems but comes with a formidable theoretical and computational challenge. In these processes, the nuclei motion cannot be confined to a single Born-Oppenheimer (BO) potential-energy surface because electronic transitions and excitations become important, and the excited carriers in turn exert a back-reaction on the lattice thus necessitating the treatment beyond the BO approximation \cite{baer2006beyond,tully2012perspective,scherrer2017mass}.  Such phenomena are exemplified by photovoltaic energy conversion \cite{ long2017nonadiabatic, cigrang2025roadmap}, photoionization \cite{evans1998photoinduced,kotur2011strong}, dissociation \cite{liu2026fly}, electron-proton transport \cite{morelli1997surface, hazra2011isotope, presiado2010excited}, and ultrafast optoelectronics \cite{ guan2022theoretical} and crack propagation \cite{theofanis2012nonadiabatic}. A fully quantum dynamical treatment of all electrons and nuclei is formally complete but becomes prohibitively expansive as the system size and propagation time increase.  This difficulty has motivated a hierarchy of nonadiabatic methods, including wave-packet propagation, surface hopping, exact factorization, and mixed quantum-classical dynamics \cite{tully1990molecular, abedi2010exact,nelson2020non}. 

The driven extended systems presents two significant computational challenges. The first is to retain the bidirectional feedback between the electronic and nuclear subsystems. Electronic excitation can drive carrier localization, lattice polarization, and structural changes  \cite{grieder2025carrier,jeschke2007microscopic}, whereas mechanical loading can generate electronic excitation and charge carriers during fracture \cite{theofanis2012nonadiabatic}. In this work, we describe this feedback loop using Ehrenfest-type mixed quantum–classical dynamics. The electronic dynamics is governed by the time-dependent Schr\"odinger equation (TDSE):
\beq
i \frac{\partial \psi}{\partial t} =  H_e^{\bar{\bfx}_{1} (t), \dots, \bar{\bfx}_{N}(t)} \psi, 
\label{eq:schodinger}
\eeq
where the electronic Hamiltonian reads
\beq
   H_e^{\bar{\bfx}_{1} (t), \dots, \bar{\bfx}_{N}(t)} = -\frac{1}{2} \sum_{k = 1}^{P}  \Delta_{\bfx_{k}} - \sum_{k = 1}^{P} \sum_{j=1}^N \frac{z_j}{| \bfx_{k} - \bar{\bfx}_{j}|} + \frac{1}{2} \sum_{k \ne m}^P \frac{1}{|\bfx_{k} - \bfx_m|}.
   \label{eq:hamiltonian}
\eeq
 Here, $\psi\equiv \psi(\bfx_1,\dots,\bfx_P,t) \in H^1(\R^{3P}\times \R^{\geq 0}; \C)$ is the time-dependent $P$-electron wavefunction (spin ignored) while the nuclear coordinates $\bar{\bfx}_1(t),\dots, \bar{\bfx}_N(t) $ are distinguished by overbars. The electronic wavefunction is antisymmetric with respect to the exchange of any two variables $\bfx_i$ and $\bfx_j$. 
 The first term on the right hand side of \eqref{eq:hamiltonian} corresponds to the kinetic energy, and the second and third terms to  Coulomb electron-nucleus and  electron-electron 
 interactions, respectively\footnote{Throughout, we use atomic units, i.e., $ m_e = |e| = \hbar = 1$, and $\frac{1}{4\pi\epsilon_0} = 1$, where $m_e$, $e$, $\hbar$, $\epsilon_0$ are the electronic mass , charge, Planck constant, and the dielectric permittivity of vacuum, respectively.}. The trajectory of a nucleus $\bar{\bfx}_i(t)$ with mass $m_i$ and atomic number $z_i$ evolves according to Newtonian dynamics on the explicit time-dependent (TD) potential energy surface $W(\bar{\bfx}_1(t),\dots,\bar{\bfx}_N(t),t)$:
 \beq
\bff_i = m_i \frac{d^2 \bar{\bfx}_{i}}{dt^2} (t) = - \nabla_{\bar{\bfx}_{i}} W(\bar{\bfx}_{1}(t),\dots, \bar{\bfx}_{N}(t), \,t),
\label{eq:MDeqn_time}
\eeq
where
 \beq
 W(\bar{\bfx}_1(t),\dots, \bar{\bfx}_{N}(t),t)  =  - \sum_{i} \int_{\R^3} \frac{z_i \rho(\bfx,t)}{|\bfx- \bar{\bfx}_{i}|} d\bfx + \frac{1}{2} \sum_{i \ne j} \frac{z_i z_j}{|\bar{\bfx}_{i} - \bar{\bfx}_{j}|},
\label{eq:tdpot}
 \eeq
 and 
 \beq
 \rho(\bfx,t)  =  P \int_{\R^{3(P - 1)}} |\psi|^2 ( \bfx,  \bfx_{2},  \dots, \bfx_{P},t) d\bfx_{2} \dots d\bfx_{P}
 \label{eq:def_rho}
 \eeq
 is the time-dependent electronic density. The electron-nuclear interaction term appearing in $W$ is known as Hellman-Feynman potential and its expression is related to the Ehrenfest theorem \cite{bornemann1996quantum,landau2013quantum}.  In short, nonadiabatic quantum 
 dynamics is the full Schr\"odinger equation for the electrons coupled to molecular dynamics for the nuclei.
 For general background on NAQD as treated here and
 its relation to other models, see
(\cite{lebris_lions}, Section 5, and \cite{bornemann1996quantum}).  A broad view of 
quantum/classical models is given by Stock and Thoss \cite{stock2005classical}.

The second challenge is to represent the spatially extended matter with the finite computational domain. For macroscopically static configurations, conventional periodic boundary conditions (PBC) accomplish this reduction by exploiting the translation invariance of the potential energy of the system. Their usual form, however, is not designed for system undergoing dynamic deformation such as extension, compression, Couette flow or vortical flow. Related questions have been studied for quantum particles subject to prescribed moving confinement such as expanding and contracting one-dimensional wells, nonlinear condensates in time-varying boxes, and radially deforming disks \cite{band2002adiabaticity,beauchard2008controllability,bisognin2013asymptotic,makowski1992behaviour,moyano2017controllability,pinder1990contracting,rouchon2003control}. More general analyses map the Schrödinger equation from a moving physical domain to a fixed reference domain through a time-dependent transformation, which introduces additional geometric or gauge-like terms into the quantum generator\cite{bernardi1997abstract,dimartino2015quantum,knobloch2015problems,duca2021schrodinger, rydin2021high}.  They nevertheless concern a particle subject to physical confinement or work with the model equations and do not provide a reduction of an unbounded and exact coupled electron–nuclear dynamics. The arbitrary time-dependent periodicity may not be dynamically consistent, that is, the non-simulated nuclei and electrons may fail to satisfy the Newton  and the Schr\"odinger equations, respectively. Moreover, the electron-mediated forces throughout the non-simulated system may fail to transform correctly. The central issue is therefore to identify an invariant manifold of the coupled electron–nuclear equations on which the degrees of freedom outside the simulated region are determined by symmetry while satisfying exactly the same equations of motion as those inside it.

Objective molecular dynamics (OMD) provides the invariant structure for the classical atomistic systems. \cite{dayal2010nonequilibrium,dayal2012design,dumitricua2007objective,pahlani2023objectivea,pahlani2023objectiveb,pahlani2023constitutive}. It depends on more general invariance of molecular dynamics that includes orthogonal transformations and (affine) time-dependent translations of the atomic positions. The underlying theorem of OMD leverages the special structure of  the MD equations arising from the two fundamental invariances of the ground state potential energy, $\varphi$. The first is frame-indifference, $\varphi(\bfQ\bar{\bfx}_1+\bfc, \bfQ\bar{\bfx}_2+\bfc, \dots) = \varphi(\bar{\bfx}_1,\bar{\bfx}_2,\dots)$, for all $\bar{\bfx}_1, \bar{\bfx}_2, \dots$, for all $\bfc$, and for all $\bfQ \in$ O(3), where O(3) is the full orthogonal group (rotations and mirror reflections) in three dimensions (3D). The second is invariance under permutations of atoms of like species: $\varphi(\bar{\bfx}_{\Pi(1)},\bar{\bfx}_{\Pi(2)},\dots) = \varphi(\bar{\bfx}_1,\bar{\bfx}_2,\dots)$, where $\Pi$ is any permutation that preserves species (atomic mass and number) \cite{dayal2010nonequilibrium}. Consequently, the invariant manifold\footnote{The invariant manifold is a surface in phase space of momentum, position, and time with the property: for a given initial conditions on the manifold at $t=0$, the solution remains on the manifold for $t>0$. The invariant manifold corresponding to OMD is independent of the material being simulated.} of the OMD equations is much bigger than the one utilized in periodic MD, offering a broader framework for simulation of time-dependent flows. 

\begin{figure}[!htbp]
    \centering
    \subfloat{\includegraphics[scale =0.82,trim={0cm 0cm 0cm 0.cm},clip]{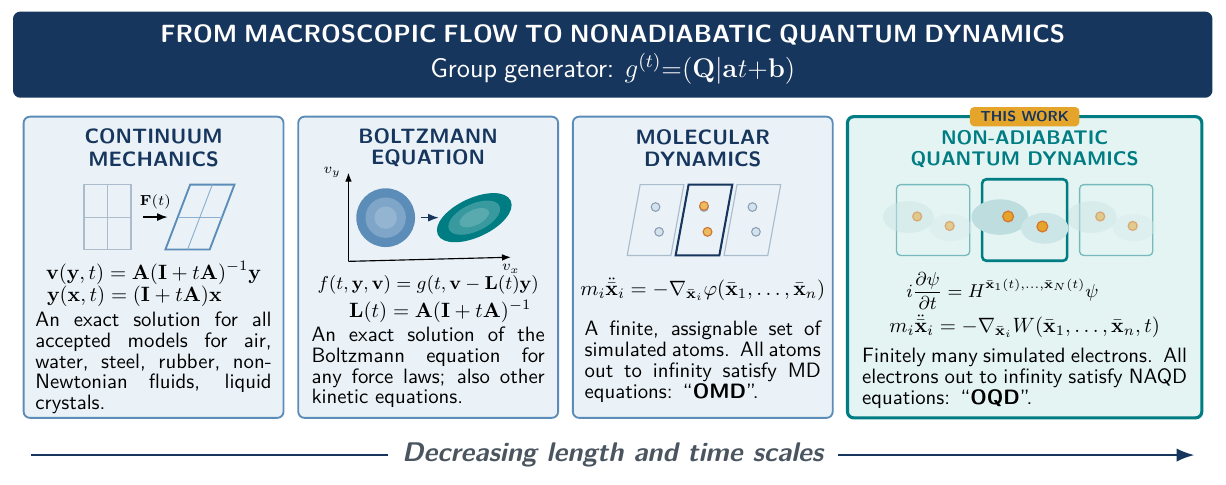}\label{fig:scale}}
\caption{Schematic showing that the invariant manifold of OMD persists across  length and time scales. Starting from the continuum scale where derformation gradient is $\bfF(t) = \bfI + t\bfA$ and Eulerian velocity field $\bfv(\bfy,t) = \bfA(\bfI+t\bfA)^{-1}\bfy$ gives the exact solution for the vast majority of accepted models in continuum mechanics. In the mesoscopic regime,  the special symmetry of OMD leads to an exact reduction of Boltzmann equation and other kinetic equations \cite{james2024kinetic}. The present work extends this invariant structure to coupled nonadiabatic quantum dynamics.}
\label{fig:time_scale}
\end{figure}

In more detail suppose the positions of the simulated atoms are denoted by $\bar{\bfx}_k(t)$, $k = 1,\dots,M \; (M>1)$ with some arbitrary initial conditions on the simulated atoms. The trajectories of the non-simulated atoms (infinite) are the orbit of a finite set of simulated atoms under a discrete group of Euclidean isometries, $G$, with special dependence on time (explained in section \ref{sec:tdig}),
\beq
\bar{\bfx}_{j,k}(t) = g_j^{(t)}(\bar{\bfx}_k(t)), \quad g_j^{(t)} \in G, \quad j = 1, 2, 3,...,
\label{eq:non-sim}
\eeq
Taking $j = 1$, e.g., $g_1^{(t)} = id$, we get the simulated atoms $\bar{\bfx}_k(t)= \bar{\bfx}_{1,k}(t)$. More precisely, even though the equations of MD are not solved for non-simulated atoms (rather, they are given by the formula \eqref{eq:non-sim}), they also satisfy the MD equations for their forces. The time dependence of the isometry group parameters provides the framework to investigate far-from-equilibrium situations. 
By choosing this group as a
translation group, OMD is able to simulate at molecular level macroscopic Eulerian velocity fields
$\bfv(\bfy, t) = \bfA(\bfI + t \bfA)^{-1}\bfy$
for an assigned $3 \times 3$ matrix $\bfA$ which include various compressible and incompressible flows and some flows with time-dependent vorticity.  Certain  flows with strong singularities, i.e., when $\bfA$ is chosen so that $\text{det}(\bfI+t \bfA)\to 0$ as  $t \to t^\star$ \cite{pahlani2023objective}, can also be simulated up to a time near $t^*$. OMD has consequently enabled finite atomistic simulations of far-from-equilibrium processes ranging from nanotube failure to reacting high-speed flows and dislocation motion \cite{dayal2010nonequilibrium,pahlani2023objectivea,pahlani2023objectiveb,pahlani2023constitutive}. OMD is not
a general purpose numerical method, but rather it produces flows that are 
dynamical solutions of the equations of motion for a variety of materials at a variety of scales (third panel in FIG.~\ref{fig:time_scale}).

 By examining in detail the statistics of a generic OMD simulation, i.e., the
 distribution of velocities of atoms in a small neighborhood of each of two atoms related by a group transformation gives a restriction on the molecular density function $f(t,\bfy,\bfv)$ of the kinetic theory of gases.  Specifically,  this restriction is $f(t,\bfy,\bfv) = g(t,\bfv-\bfA(\bfI+t\bfA)^{-1}\bfy)$.  Substituting this anstaz into the Maxwell-Boltzmann equation leads to an exact reduction and the Maxwell-Boltzmann equation inherits the invariant manifold of OMD exactly \cite{james2024kinetic}.  These results
 extend to other kinetic equations obtained under
 different scalings \cite{james2024kinetic} (second panel in Fig.~\ref{fig:time_scale}). 

At the continuum scale, for every widely accepted model of fluids or solids in
continuum mechanics, in the absence of body forces, the OMD velocity field,
$\bfv(\bfy, t) = \bfA(\bfI + t \bfA)^{-1}\bfy$,
is an exact solution of the equations of
motion for every choice of $\bfA$ as illustrated first panel in Fig.~\ref{fig:time_scale}. Besides conventional continuum models like the Navier-Stokes equations or nonlinear elasticity,  this also includes less traditional models such as  non-Newtonian fluids or the Ericksen-Leslie theory of liquid crystals (with an appropriate transformation law
for the director). If thermodynamics is included in the continuum theory, OMD does not
give a universal macroscopic  law for the evolution of energy or entropy. 
In typical cases, once $\bfv(\bfy, t) = \bfA(\bfI + t \bfA)^{-1}\bfy$ is inserted into the equations of motion of the continuum theory, the energy equation becomes an (material dependent) ordinary differential equation for the temperature.

In the static case  this includes symmetry-adapted Kohn-Sham Density Functional Theory (KS-DFT) \cite{banerjee2021ab,ghosh2019symmetry}. Due to this development, various remarkable features including electronic phase transformation due to torsional deformation \cite{yu2022density}, strongly correlated behavior \cite{yu2024carbon}, and topological edge states \cite{sharma2025electronic} in variety of nanotubes were efficiently calculated. These developments, however, do not establish whether a NAQD remains compatible with the evolving atomistic symmetry. Extending the exact invariant-manifold of OMD to NAQD is the objective of the present work, shown in fourth panel in Fig.~\ref{fig:time_scale}.

In this paper, we introduce objective non-adiabatic quantum dynamics, or briefly, Objective Quantum Dynamics (OQD), governed by \ref{eq:schodinger}-- \eqref{eq:def_rho} and with group invariance generalized from \eqref{eq:non-sim}.  Here, only a subset of electrons (the ``simulated electrons") and nuclei (the ``simulated nuclei") are treated using \eqref{eq:schodinger}-\eqref{eq:def_rho}, while the remaining nuclei and electrons are obtained by acting a time-dependent isometry group on the simulated electrons (described by the wavefunction) and nuclei.  The main issues are: 1) determining the action of the group on the many-body wavefunction  and 2) dealing with the explicit time dependence of $W$, which at the outset seems problematic, and 3) quantifying the interaction of the nonsimulated electrons on the simulated electrons. The conclusion is completely analogous to that of OMD:  the main Theorem \ref{ell} below says that all nuclei and electrons, simulated and non-simulated, satisfy the equations of OQD, \eqref{eq:schodinger}-\eqref{eq:def_rho}. This invariant manifold is also inherited by Time-dependent Hartree-Fock (TDHF) equations obtained by taking Slater-determinant of many-body functions.  (Theorem \ref{pro:pro_tdhfg}). 


 We choose higher order finite-difference method in combination with self-consistent midpoint Crank–Nicolson method and velocity-Varlet algorithm for discretizing and solving the governing equations in this work. To avoid the spurious electronic effects due to continuous interpolation in the deforming fundamental domain, we transformed the equations in Lagrangian coordinate, in the fixed reference domain. This gives us a working realization of an first principle computational tool, OQDFlow, that can be used to perform  quantum dynamics simulations of system undergoing affine time-dependent deformation in a systematic and efficient manner. As a proof of principle, we study an infinite system of warm dense Hydrogen fluid. While the hydrogen system is simple and computationally tractable, it embodies many central themes of modern condensed-matter physics and astrophysics. \cite{motta2020ground, morales2010evidence}.

The rest of the paper is organized as follows. In section \ref{sec:tdig}, we give a brief overview on the isometry groups where the group elements have affine time-dependence. Section \ref{oqd} presents objective quantum dynamics, where we give transformation laws for 1) nuclear-nuclear and nuclear-electrons forces (subsection \ref{sec:TNN}), 2) nuclear trajectories (subsection \ref{sec:ODENM}), and 3) for the TDSE for electron dynamics (subsection \ref{sec:TLSC}).  In section \ref{sec:tdhf}, we show that the time-dependent Hartree-Fock (TDHF) equations also satisfies the theorems of OQD.  We then discuss the numerical implementation and approximations taken to solve the objective TDHF coupled with OMD in section \ref{sec:alg}. We present the application of our method in section \ref{sec:application} and we conclude in section \ref{sec:conclusions} with possible future directions.


\section{Time-dependent isometry groups}\label{sec:tdig}
OQD relies on the discrete group of isometries as
defined below:
\begin{definition}
    An isometry group $G$ in 3D consists of $N \le \infty$ elements of the form $g = (\bfQ|\bfc)$, $\bfQ\in$ {\rm O(3)}, and $\bfc \in \R^3$. Its action on $\R^3$ is   $g(\bfx) = \bfQ\bfx+\bfc$. The group $G$ is  discrete, i.e., $G(\bfx) = \{g_1(\bfx),g_2(\bfx),\dots\, g_N(\bfx)\}$ has no accumulation point for any $\bfx \in \R^3$. 
The product, identity and inverse of isometries are
defined by:
\beqs
g_1g_2(\bfx) &:=& g_1(g_2(\bfx))\ \ \forall \bfx \in  \mathbb R^3 \nonumber\\
\implies & &(\bfQ_1|\bfc_1)(\bfQ_2|\bfc_2) = (\bfQ_1\bfQ_2|\bfQ_1\bfc_2+\bfc_1), \nonumber \\
g_1 &=& id = (\bfI|\bf0), \nonumber \\
g^{-1} &=& (\bfQ|\bfc)^{-1} = (\bfQ^{T}|-\bfQ^{T}\bfc).
\eeqs
\end{definition}

\noindent  $N \in \mathbb Z$ and infinite ($N  = \infty$)
groups such as translation groups are allowed.  

The International Tables of Crystallography \cite{hahn1983international} contains a listing of 
isometry groups.  This includes the various 3D translation groups of the form $G_T = \{t^p_1t^q_2t^r_3:t_i =  (\bfI|\bfe_i), p,q,r \in \mathbb Z\}$, where $\bfe_1, \bfe_2$ and $\bfe_3$ are the three linearly independent vectors. Also allowed in OMD and OQD are the subperiodic isometry groups, such as ring groups, helical groups, rod groups and net groups.

Similar to OMD, the nuclear position vectors are labeled by double index $(i,k)$: $\bar{\bfx}_{i,k}, i = \{1,\dots, N\}, k = \{1,\dots, M\}$. For infinite groups like translation groups, $N = \infty$, while $M$ is always finite. Nuclei with positions $\bar{\bfx}_{i,k}$ and $\bar{\bfx}_{j,k}$ must be the same element, i.e., have the same atomic mass and number. 

Both OMD and OQD allow only a rather restricted time dependence of the group:  for each group element $g_i^{(t)} = (\bfQ_i|\bfc_i(t)) \in G^{(t)}$,
$\bfQ_i \in$ O(3)  must be constant and the translation $\bfc_i(t)$ must be affine in time:
\beq
\bfc_i(t) = \bfa_i t+\bfb_i, \quad \bfa_i,\bfb_i \in \R^3, \quad i = 1,\dots,N.
\label{eq:ctimedep}
\eeq
The $\bfa_i$ and $\bfb_i$, $i = 1, \dots, N$, have to be chosen to respect the group properties. The
fact that $\bfa_i$ and $\bfb_i$ depend on $i$ is
responsible for the variety of non-Galilean macroscopic motions
achievable with the method.
The $\bar{\bfx}_{j,k}(t), \ j \in \{1, \dots, N\}, k \in 
\{1,\dots, M\}$, denote
positions of the nuclei.  Some nuclei 
denoted by
$\bar{\bfx}_{k}(t) := \bar{\bfx}_{1,k}(t)$ are designated as {\it simulated nuclei}.
These restrictions imply that
the positions $\bar{\bfx}_{j,k}(t)$ of the nuclei 
satisfy
 \beq
 \frac{d^2\bar{\bfx}_{j,k}(t)}{dt^2} = \frac{d^2}{dt^2}g_j^{(t)}(\bar{\bfx}_{k}(t)) = \bfQ_j\frac{d^2\bar{\bfx}_k(t)}{dt^2},
 \label{eq:nonsimcond}
 \eeq
 where, the relabeled group elements,  $g_j^{(t)} = (\bfQ_j|\bfa_j t+\bfb_j)\in G^{(t)}, \ j = \{1, 2,\dots, N\}$, exhibit the allowed time dependence\footnote{For more information on this restriction see \cite{dayal2010nonequilibrium}, Section 4.}.  Below, we often use the terminology
 ``nucleus $(j,k)$'' to denote the nucleus at position $\bar{\bfx}_{j,k}$.
 
 Note that, in OQD (also OMD), although each group element 
 by itself is a Galilean transformation, the full system of simulated and nonsimulated  nuclei do not undergo a Galilean transformation, except in the trivial case of one simulated atom. A similar statement holds in OQD. 
In fact, using a translation group with as few as two simulated atoms
in 2D, the differential equations of OMD with a translation group already evidently represent a chaotic dynamical system.  The derivation below is not restricted
to  translation groups.

\section{Objective quantum dynamics}
\label{oqd}
This section gives a formal treatment
of the equations of molecular dynamics \eqref{eq:MDeqn_time} for the nuclear positions. The
right hand side of this
equation involves 
repulsive nuclear-nuclear and attractive nuclear-electron interactions. Each of
these terms by itself is badly behaved
in the case of an infinite group like
the translation group.  If these two
contributions are combined, the situation is expected to be better due to screening, but still the
sum of nuclear-nuclear and nuclear-electron terms is typically infinite
for infinite groups.  This situation 
already occurs in OMD: the typical potential
energy of an infinite system of atoms is
infinite.  However, its derivative, i.e., the force on an atom, is typically
well behaved.  Hence, in our formal treatment  we deal directly with the force on a nucleus here.

Based on group invariance of the electronic density \eqref{inv_den} and \eqref{eq:non-sim}, we observe the following consequences on the potential energy defined in \eqref{eq:tdpot}.

\subsection{Transformation laws for the sum of Nuclear-Nuclear (NN) and Nuclear-Electron (NE) forces} \label{sec:TNN}

In the OQD framework, it will follow from properties of the wavefunction that the electronic density defined in \eqref{eq:def_rho} is invariant under the the given time-dependent isometry group, i.e., for all $\bfx\in \R^3$, 
\beq
\rho(\bfx,t)  = \rho(g_i^{(t)}(\bfx), t), \quad i = 1, \dots, N, \quad  t > 0.  \label{inv_den}
\eeq
For expected regularity of
the electronic density and for infinite groups such as the translation group, integrals of the
form 
\beq
\int_{\R^3} \frac{z_k \rho(\bfx,t)}{|\bfx - \bar{\bfx}_{i,k}(t)|} d\bfx 
\eeq
blow up due to the behavior of the integrand as $\bfx \to \infty$. Thus, the formal  definition of the potential energy \eqref{eq:tdpot} is
typically unbounded for infinite groups like translation groups, but the derivative is expected to be well-defined for charge neutral systems of nuclei and electrons, as well as for some systems with modest charge imbalance.  We therefore formally
consider the force on the nucleus at
position $\bar{\bfx}_i$ due to other nuclei and electrons:
\begin{eqnarray}
\label{eq:rhs}
\lefteqn{- \nabla_{\bar{\bfx}_{i,k}} W(\bar{\bfx}_{1,1}(t),\dots, \bar{\bfx}_{N,M}(t), \,t) } \\
&=& - \nabla_{\bar{\bfx}_{i,k}} \left(-\sum_{i,k}\int_{\R^3} \frac{z_k \rho(\bfx,t)}{|\bfx - \bar{\bfx}_{i,k}(t)|}d\bfx + \frac{1}{2} 
\sum_{\substack{(j,\ell), (i,k) \\ (j,\ell) \neq  (i,k)}} \frac{z_k z_{\ell}}{|\bar{\bfx}_{j,\ell} - \bar{\bfx}_{i,k}|} \right) \nonumber\\
&=&  \int_{\R^3} \frac{z_k\, \rho(\bfx,t)(\bfx - \bar{\bfx}_{i,k}(t))}{|\bfx - \bar{\bfx}_{i,k}(t)|^3}d\bfx - \sum_{\substack{(j,\ell) \\ (j,\ell) \neq  (i,k)}} \frac{z_k z_{\ell}(\bar{\bfx}_{j,\ell} - \bar{\bfx}_{i,k})}{|\bar{\bfx}_{j,\ell}-\bar{\bfx}_{i,k}|^3}. \nonumber
\end{eqnarray}
It can be seen from \eqref{eq:rhs} that, if 
the electronic density is arranged in a suitable
approximately charge-neutral way around the ions,
then \eqref{eq:rhs} is well defined with  the integral treated as a singular integral and the
sum having a corresponding cut-off.
\subsection{Ordinary differential equations for
the simulated nuclei} \label{sec:ODENM}
As in OMD, we assume the motions $\bar{\bfx}_{j,k}(t)$  of the nuclei are determined from the motions
of the simulated nuclei by the instantaneous group action,
\beq
\bar{\bfx}_{i,k}(t) = g_i^{(t)}(\bar{\bfx}_{1,k}(t)),
\quad j = 1, \dots , N, \ \ k = 1, \dots, M.
\label{eq:non-sim1}
\eeq
from which it follows from the allowed time dependence of the group that
 \beq
 \frac{d^2\bar{\bfx}_{i,k}(t)}{dt^2} = \frac{d^2}{dt^2}g_i^{(t)}(\bar{\bfx}_{1,k}(t)) = \bfQ_i\frac{d^2\bar{\bfx}_{1,k}(t)}{dt^2}.
 \label{eq:nonsim2}
 \eeq
(Recall that $M$ is finite.) 

The goal now is to show that, like
\eqref{eq:nonsim2}, the force on ion $(j,k)$
is expressible as $\bfQ_i \times $(force on the
simulated ion $(1,k)$), thus reducing the 
dynamical equations for all the ions to a finite system of ordinary
differential equations for the simulated ions.
The motions of the non-simulated ions can then 
be recovered by simply applying the appropriate group element to the simulated atoms using \eqref{eq:non-sim1}.

While it is the sum of the two terms in \eqref{eq:nonsim2} that 
is nominally well-defined, for clarity we formally consider the two terms in \eqref{eq:rhs} separately. Beginning with the first term of 
\eqref{eq:rhs}, we introduce \eqref{eq:non-sim1} 
and make the change of variables $\bfx = g_i^{(t)}(\bfy)$:
\begin{dmath}
 \int_{\R^3} \frac{z_k\, \rho(g_i^{(t)}(\bfy),t)(g_i^{(t)}(\bfy) - g_i^{(t)}(\bar{\bfx}_{1,k}(t)))}{|g_i^{(t)}(\bfy) - g_i^{(t)}(\bar{\bfx}_{1,k}(t))|^3}d\bfy \\
 =  \bfQ_i \int_{\R^3} \frac{z_k\, \rho(\bfy,t)(\bfy - \bar{\bfx}_{1,k}(t))}{|\bfy - \bar{\bfx}_{1,k}(t)|^3}d\bfy. 
 \label{eq:nn1}
\end{dmath}
Similarly, the second term in \eqref{eq:rhs} can be
written (note $i,k$ are fixed)
\beqs
-\sum_{j,\ell \ne i,k} \frac{z_k z_{\ell}(\bar{\bfx}_{j,\ell} - \bar{\bfx}_{i,k})}{|\bar{\bfx}_{j,\ell} - \bar{\bfx}_{i,k}|^3}
&=& -\bfQ_i\sum_{p,\ell \ne 1,k} \frac{z_k z_{\ell}\,
(\bar{\bfx}_{p,\ell} - \bar{\bfx}_{1,k})}{|\bar{\bfx}_{p,\ell} - \bar{\bfx}_{1,k}|^3}. 
\label{eq:nn2}
\eeqs
To arrive at \eqref{eq:nn2}, we write
\beqs
\bar{\bfx}_{i,k} &=& g_{i}^{(t)}(\bar{\bfx}_{1,k}) 
= \bfQ_i \bar{\bfx}_{1,k} + \bfa_i t + \bfb_i,\\
\bar{\bfx}_{j,\ell} &=& g_{i}^{(t)}g_{i}^{(t)-1}g_j^{(t)}(\bar{\bfx}_{1,\ell}) 
= \bfQ_i g_{i}^{(t) -1} g_j^{(t)}(\bar{\bfx}_{1,\ell}) + \bfa_i t + \bfb_i,
\eeqs
so that
\beq
\bar{\bfx}_{j,\ell} - \bar{\bfx}_{i,k}
= \bfQ_i[g_{i}^{(t)-1}g_j^{(t)}(\bar{\bfx}_{1,\ell}) - \bar{\bfx}_{1,k}].
\eeq
Then we notice that $g_{i}^{(t)-1}g_j^{(t)}$
summed over $j = 1, \dots, N$ is the same as
$g_p^{(t)}$ summed from $p = 1, \dots, N$.

Summarizing, we have shown that, formally, the 
equations for the nuclei become ordinary
differential equations for the simulated  nuclei.
That is, combining \eqref{eq:rhs}, \eqref{eq:nonsimcond} and \eqref{eq:nn2} we have
reduced the dynamical equations for all the nuclei
to the following equations for the simulated
nuclei:
\begin{eqnarray}
\label{eq:odes_for_ions}
\lefteqn{\frac{d^2 \bar{\bfx}_{1,k}}{dt^2} = } \\
& & \int_{\R^3} \frac{z_k\, \rho(\bfy,t)(\bfy - \bar{\bfx}_{1,k}(t))}{|\bfy - \bar{\bfx}_{1,k}(t)|^3}d\bfy - \sum_{\substack{(p.\ell) \\(p,\ell) \ne (1,k)}} \frac{z_k z_{\ell}\,
(g_p^{(t)}\bar{\bfx}_{1,\ell} - \bar{\bfx}_{1,k})}{|g_p^{(t)}\bar{\bfx}_{1,\ell} - \bar{\bfx}_{1,k}|^3}. \nonumber
\end{eqnarray}
Once the motions of the simulated nuclei are obtained from \eqref{eq:odes_for_ions}, then 
the positions of all the other nuclei are updated by applying
the group to the simulated nuclei.  Since the
two terms on the right hand side compete effectively at the scale of individual molecules, screening will play a useful role for the evaluation of \eqref{eq:odes_for_ions}, especially in cases that there is macroscopic charge neutrality.
\subsection{Transformation law for solutions of Schr\"odinger equation for the electrons}
\label{sec:TLSC}
 Now we  examine the Schr\"odinger equation for the evolution of electrons. We first
 identify some invariance properties of solutions
 for the electronic density under the condition that the nuclear positions
 are determined by the group acting on the
 simulated nuclei.

 The first step is to identify the group transform of the wavefunction and to show that it satisfies
 the Schr\"odinger equation. This transform involves a time-dependent prefactor \eqref{gtpsi} that depends on the group parameters.

\begin{proposition}[Invariance of TDSE under time dependent group] \label{ell} Let the wavefunction $\psi(\bfx_1,  \dots, \bfx_P, t)$ satisfy 
\beq
 i \frac{\partial \psi}{\partial t} =  H_e^{\bar{\bfx}_{1,1} (t), \dots, \bar{\bfx}_{N,M}(t)} \psi
 \label{see}
 \eeq
 
 \begin{dmath*}
   H_e^{\bar{\bfx}_{1,1} (t), \dots, \bar{\bfx}_{N,M}(t)} = -\frac{1}{2} \sum_{k = 1}^{P}  \Delta_{\bfx_{k}} - \sum_{k = 1}^{P} \sum_{(i,j)} \frac{z_j}{| \bfx_{k} - \bar{\bfx}_{i,j}|} + \frac{1}{2} \sum_{k \ne m}^P \frac{1}{|\bfx_{k} - \bfx_m|}, 
   \label{elect1}
\end{dmath*}
with nuclear motions determined by the simulated nuclei:
\beq
\bar{\bfx}_{i,j}(t) = g_i^{(t)}(\bar{\bfx}_{1,j}(t)), \quad i = 1, \dots, N, \quad j = 1, \dots,M, \quad t > 0.  \label{inv_nuc1}
\eeq
Here, as above, $G^{(t)} = \{ g_1^{(t)}, g_2^{(t)}, \dots, g_N^{(t)} \}$ is a discrete group of isometries with affine time dependence of the translations.  Fix $1 \le \ell \le N$, where 
$g_{\ell}^{(t)}(\bfx)= \bfQ_{\ell} \bfx + \bfa_{\ell} t + 
\bfb_{\ell} \in G^{(t)} $, $\bfQ \in $ \rm{O}(3), $\bfa_{\ell}, \bfb_{\ell} \in \R^3$, {\it and define the group transform of a solution by}
\beqs
& & \  \ \psi_{\ell}(\bfx_1,  \dots, \bfx_P, t) \nonumber\\
\ \ \ & & \, \hspace{2cm}= e^{-i \bfa_{\ell} \cdot (\bfQ_{\ell}(\bfx_1 + \cdots + \bfx_P) + \frac{1}{2}P \bfa_{\ell} t + \frac{1}{2} P \bfb_{\ell}    ) }\psi(g_{\ell}^{(t)}(\bfx_1),  g_{\ell}^{(t)}(\bfx_2),  \dots, g_{\ell}^{(t)}(\bfx_P), t)  \label{gtpsi}
\eeqs
{\it Then $\psi_{\ell}(\bfx_1,   \dots, \bfx_P, t)$
formally satisfies \eqref{see} for each $\ell = 1, \dots, N$.}
\end{proposition}
\begin{proof} The proof follows by direct calculation of the terms in  the Schr\"odinger equation.  Using  \eqref{gtpsi}, we get 
\beqs
\label{eq:timederi}
i \frac{\partial\psi_\ell}{\partial t} &=& e^{-i \bfa_\ell \cdot (\bfQ_\ell(\bfx_1 + \cdots + \bfx_P) + \frac{1}{2}P \bfa_\ell t  + \frac{1}{2} P \bfb_{\ell} ) }
\left(\frac{1}{2} P|\bfa_\ell|^2 \psi + i \bfa_\ell \cdot \sum_{k=1}^P\psi_{\bfx_k} + i \frac{\partial\psi}{\partial t} \right), \\
\ \  \frac{1}{2} \psi_{\ell \bfx_k} &=& \frac{1}{2} e^{-i \bfa_\ell \cdot (\bfQ_\ell(\bfx_1 + \cdots + \bfx_P) + \frac{1}{2}P \bfa_\ell t + \frac{1}{2} P \bfb_{\ell} )} \big(-i \bfQ_{\ell}^T \bfa_\ell \psi  + \bfQ_{\ell}^T\psi_{\bfx_k}\big). \nonumber
 \eeqs
 Differentiating, the latter, we calculate the kinetic energy,
\beqs
\label{eq:kinenergy}
 \lefteqn{ \hspace{15mm} -\frac{1}{2} \sum_{k=1}^P \psi_{\ell \bfx_k \bfx_k}} \\
&=& 
\frac{1}{2} e^{-i \bfa_\ell \cdot (\bfQ_\ell(\bfx_1 + \cdots + \bfx_P) + \frac{1}{2}P \bfa_\ell t + \frac{1}{2} P \bfb_{\ell})} \bigg(P|\bfa_\ell|^2 \psi +  
 2i \bfQ_\ell^T\bfa_\ell\cdot \bfQ_\ell^T\sum_{k=1}^P\psi_{\bfx_k} - {\rm tr} \big(\bfQ_\ell^T\sum_{k=1}^P\psi_{\bfx_k \bfx_k}\bfQ_\ell\big) \bigg). \nonumber \\
&=& e^{-i \bfa_\ell \cdot (\bfQ_\ell(\bfx_1 + \cdots + \bfx_P) + \frac{1}{2}P \bfa_\ell t + \frac{1}{2} P \bfb_{\ell})} \bigg(\frac{1}{2} P|\bfa_\ell|^2 \psi + i \bfa_\ell\cdot \sum_{k=1}^P\psi_{\bfx_k} - \frac{1}{2} \sum_{k=1}^P\psi_{\bfx_k \bfx_k}\bigg). 
\nonumber
\eeqs
Combining these results,
\beqs
\label{2terms}
\lefteqn{i \frac{\partial }{\partial t}\psi_\ell(\bfx_1,\dots,\bfx_P,t) + \frac{1}{2} \sum_{k=1}^P \Delta_{\bfx_k}\psi_{\ell}(\bfx_1,\dots,\bfx_P,t)}  \\
&=& e^{-i \bfa_\ell \cdot (\bfQ_\ell(\bfx_1 + \cdots + \bfx_P) + \frac{1}{2}P \bfa_\ell t + \frac{1}{2}P \bfb_\ell) }
\bigg(  i \frac{\partial }{\partial t}\psi(g_{\ell}^{(t)}(\bfx_1),  \dots, g_{\ell}^{(t)}(\bfx_P), t)  \nonumber\\
& & \quad + \quad\frac{1}{2}\sum_{k=1}^P\Delta_{\bfx_k}\psi(g_{\ell}^{(t)}(\bfx_1),  \dots, g_{\ell}^{(t)}(\bfx_P), t)  \bigg).  \nonumber
\eeqs
Finally, we consider the potential part of the electronic Hamiltonian:
\beqs
V\psi_{\ell}
&=&  - \sum_{k = 1}^{P} \sum_{(i,j)} \frac{z_j \psi_{\ell}(\bfx_1, \dots, \bfx_P,t)}{| \bfx_{k} - \bar{\bfx}_{i,j}(t)|} + \frac{1}{2} \sum_{k \ne m}^P \frac{\psi_{\ell}(\bfx_1, \dots, \bfx_P,t)}{|\bfx_{k} - \bfx_m|} \\
&=& e^{-i \bfa_{\ell} \cdot (\bfQ_{\ell}(\bfx_1 + \cdots + \bfx_P) + \frac{1}{2}P \bfa_{\ell} t +
\frac{1}{2}P \bfb_{\ell}) } \bigg( - \sum_{k = 1}^{P} \sum_{(i,j)} \frac{z_j \psi(g_{\ell}^{(t)}(\bfx_1), \dots, g_{\ell}^{(t)}(\bfx_P), t) }{|\bfx_k - g_i^{(t)}(\bar{\bfx}_{1,j}(t))|} \nonumber \\
& &\quad \quad +\quad  \frac{1}{2} \sum_{k \ne m}^P \frac{\psi(g_{\ell}^{(t)}(\bfx_1), \dots, g_{\ell}^{(t)}(\bfx_P), t)}{|\bfx_k - \bfx_m|} \bigg)  \nonumber \\
&=& e^{-i \bfa_{\ell} \cdot (\bfQ_{\ell}(\bfx_1 + \cdots + \bfx_P) + \frac{1}{2}P \bfa_{\ell} t +
\frac{1}{2}P \bfb_{\ell}) } \bigg( - \sum_{k = 1}^{P} \sum_{(i,j)} \frac{z_j \psi(g_{\ell}^{(t)}(\bfx_1), \dots, g_{\ell}^{(t)}(\bfx_P), t) }{|g_{\ell}^{(t)}(\bfx_k) - g_{\ell}^{(t)}g_i^{(t)}(\bar{\bfx}_{1,j}(t))|} \nonumber \\
& &\quad \quad +\quad  \frac{1}{2} \sum_{k \ne m}^P \frac{\psi(g_{\ell}^{(t)}(\bfx_1), \dots, g_{\ell}^{(t)}(\bfx_P), t)}{|g_{\ell}^{(t)}(\bfx_k) - g_{\ell}^{(t)}(\
\bfx_m)|} \bigg)  \nonumber \\
&=& e^{-i \bfa_{\ell} \cdot (\bfQ_{\ell}(\bfx_1 + \cdots + \bfx_P) + \frac{1}{2}P \bfa_{\ell} t
+\frac{1}{2}P \bfb_{\ell}) } \bigg( - \sum_{k = 1}^{P} \sum_{(i',j)} \frac{z_j \psi(g_{\ell}^{(t)}(\bfx_1), \dots, g_{\ell}^{(t)}(\bfx_P), t) }{|g_{\ell}^{(t)}(\bfx_k) - g_{i'}^{(t)}(\bar{\bfx}_{1,j}(t))|} \nonumber \\
& &\quad \quad  +\quad  \frac{1}{2} \sum_{k \ne m}^P \frac{\psi(g_{\ell}^{(t)}(\bfx_1), \dots, g_{\ell}^{(t)}(\bfx_P), t)}{|g_{\ell}^{(t)}(\bfx_k) - g_{\ell}^{(t)}(\bfx_m)|} \bigg)  \nonumber \\
&=&
e^{-i \bfa_{\ell} \cdot (\bfQ_{\ell}(\bfx_1 + \cdots + \bfx_P) + \frac{1}{2}P \bfa_{\ell} t
+\frac{1}{2}P \bfb_{\ell}) }(V\psi) (g_{\ell}^{(t)}(\bfx_1), \dots, g_{\ell}^{(t)}(\bfx_P), t) \nonumber
\eeqs
Here, we have used closure property of the group, $g_{\ell}^{(t)}\calG^{(t)} = \calG^{(t)}$.
This with \eqref{2terms} completes the proof.
\end{proof}

This proposition implies that if the initial data for $\psi$ satisfies certain restrictions relating to the group, then it continues to be group invariant for $t>0$. This has the effect of reducing the domain of wave functions to a fundamental domain.

\begin{lemma}[Uniqueness of wavefunction] Suppose electronic wavefunction $\psi$ satisfies the time dependent Schr\"odinger equation
\label{lm:comp}
\beq
 i \frac{\partial \psi}{\partial t} =  H_e^{\bar{\bfx}_{1,1} (t), \dots, \bar{\bfx}_{N,M}(t)} \psi, \label{se}
\eeq
subject to initial conditions satisfying the identities
\beqs
& &\ \ \psi(\bfx_1, \dots, \bfx_P,0)  \\
& & \;=e^{-i a_{\ell} \cdot (\bfQ_{\ell} (\bfx_1 + \cdots + \bfx_P) + \frac{1}{2}P \bfb_{\ell})}\psi(g_{\ell}^{(0)}(\bfx_1), \dots g_{\ell}^{(0)}(\bfx_P), 0), \quad
1 \le \ell \le N.  \nonumber \label{iden1}
\eeqs
Then $\psi$ satisfies the identities
\begin{dmath}
\psi(\bfx_1, \dots, \bfx_P,t) = e^{-i a_{\ell} \cdot (\bfQ_{\ell}(\bfx_1 + \cdots + \bfx_P) + \frac{1}{2}P \bfa_{\ell} t +\frac{1}{2}P \bfb_{\ell}) }\psi(g_{\ell}^{(t)}(\bfx_1),  g_{\ell}^{(t)}(\bfx_2),  \dots, g_{\ell}^{(t)}(\bfx_P), t)  \label{iden2}
\end{dmath}
for $\ell = 1, \dots, N$ and $t>0$.  That is, $\psi$ is determined by its
behavior on a fundamental domain of $G^{(t)}$
for each $t>0$.
\end{lemma}

\begin{proof} Define $\psi_{\ell}$ by \eqref{gtpsi}.
Then \eqref{iden1} implies that
\beq
\psi(\bfx_1, \dots, \bfx_P,0) = \psi_{\ell}(\bfx_1, \dots, \bfx_P,0).
\eeq
Since $\psi$ and $\psi_{\ell}$ both satisfy the same Schr\"odinger equation with the same initial conditions, then by uniqueness, they are the same.  Clearly, \eqref{iden2} serves to define $\psi$ on all of space once it is known on a time-dependent fundamental domain $F\!D^{(t)}$ since, by definition of a fundamental domain, 
\beq
\bigcup_{\ell} \big(g_{\ell}^{(t)}(\bfx_1),  g_{\ell}^{(t)}(\bfx_2),  \dots, g_{\ell}^{(t)}(\bfx_P)\big)\bigg|_{(\bfx_1, \dots, \bfx_P) \in F\!D} = \R^{3P}.
\eeq
\end{proof}

\section{OQD for time-dependent Hartree-Fock}
\label{sec:tdhf}
Similar to the time-independent case, even with the simplifications provided by group invariance, the direct solution of the TDSE presents a difficult problem except in  simple cases. In this section, we demonstrate that OQD is compatible with the use of Slater determinants, leading to a group invariant Hartree-Fock method. We first briefly outline the derivation of coupled equations of time-dependent Hartree-Fock (TDHF)  and Hellmann-Feynman nuclear dynamics, and then show the invariance of TDHF under the groups treated here.

The many-body wavefunction $\psi$ is the solution of the TDSE; however, determining its explicit form becomes exponentially difficult for the systems with increasing numbers of electrons.  To make the problem tractable, common approach is to restrict the wavefunction to lie in the antisymmetric electronic Hilbert space, $\mathcal{H}_e = \bigwedge_{i=1}^P L^2(\R^3\times \R^{\geq 0},\C)$. The space can be approximated by a single Slater determinant of single electron wavefunctions $\phi_i(\bfx,t)$:
\beq \label{slatterdet1}
\mathcal{A} = \left\{ \psi (\bfx_1, \dots, \bfx_p,t) = \frac{1}{\sqrt{P!}} \det(\phi_i(\bfx_j,t)),\;  \phi_i \in H^1(\R^3\times \R^{\geq 0},\C),\;\int_{\R^3}\phi^*_i\phi_j = \delta_{ij}\right\} . \nonumber
\eeq
This guarantees the correct normalization and antisymmetry.  To obtain the TDHF equations, we write a stationary action principle, based on
\beq
S = \int_0^T\langle\psi(\bfx_1,\dots,\bfx_p,t),(i\partial_t - H_e^{\bar{\bfx}_{1,1}(t),\dots \bar{\bfx}_{N,M}(t)})\psi(\bfx_1,\dots,\bfx_p,t\rangle dt,
\label{eq:exp_eq}
\eeq
where, $\langle \cdots, \cdots\rangle$ is the  $L^2$ inner product over the electronic coordinates.  The Euler-Lagrange equations of \eqref{eq:exp_eq} give the TDHF equations:
\beq
i\frac{\partial \phi_m}{\partial t} = \bar{H}_{\psi}^{\bar{\bfx}_{1,1}(t),\dots \bar{\bfx}_{N,M}(t)} \phi_m + \sum_{m}\lambda_{mn}(t)\phi_n,
\label{eleq}
\eeq
where, $\bar{H}_\psi^{\bar{\bfx}_{1,1}(t),\dots \bar{\bfx}_{N,M}(t)}$ is the Hartree-Fock (HF) Hamiltonian,
\beq
\bar{H}_\psi^{\bar{\bfx}_{1,1}(t),\dots \bar{\bfx}_{N,M}(t)} = -\frac{1}{2}\Delta - \sum_{(i,j)}\frac{z_j}{|\cdot - \bar{\bfx}_{i,j}(t)|} + \left(\sum_{k=1}^P|\phi_k|^2\star\frac{1}{|\bfx|}\right) - \sum_{k=1}^P \left(\phi_k^* \star\frac{1}{|\bfx|}\right)\phi_k .
\label{Hartreefock}
\eeq
Here, the third and fourth terms are the Hartree and non-local Fock exchange potential ($V_{ex}$), respectively, and $\star$ represents the convolution. The Hermitian matrix $\lambda_{mn}$ should not be confused with the Lagrange multiplier used to enforce the orthonormality constraint, $\int_{\R^3} \phi^*_m(\bfx)\phi_n(\bfx) = \delta_{mn}$, in the minimization problem of time-independent Schr\"odinger equation. In time dependent situation, if the wavefunctions are orthonormal at $t=0$, then Hermiticity of HF Hamiltonian $H_{\psi}$ automatically preserves orthonormality over time, $t>0$. Rather, 
$\lambda_{mn}$ represents the degrees of freedom associated with gauge invariance, $\phi_m(\bfx,t) = U_{mn}(t)\phi_n(\bfx,t)$, of Slater wavefunctions under the unitary transformation $U(t)$. The matrix $\lambda_{mn}(t)$ represents the generator of internal rotation, which describes how the orbitals rotate. Without loss of generality, we choose $\lambda_{mn}(t) = 0$.  The TDHF equation then becomes
\beq
i\frac{\partial }{\partial t}\phi_m(\bfx,t) = \bar{H}_{\psi}^{\bar{\bfx}_{1,1}(t),\dots \bar{\bfx}_{N,M}(t)} \phi_m(\bfx,t), \; \; \bfx \in \R^3, \; t \in \R^{\geq 0}, 
\label{eleql0}
\eeq
and the electron density is
\beq
\rho_{\psi}(\bfx,t) = \sum_{m =1}^P \vert\phi_m(\bfx,t)\vert ^2.
\eeq
 This approximation reduces the original many-body TDSE to a Cauchy problem of coupled systems consisting of the TDHF equations for the electronic degrees of freedom and the Hellmann-Feynman equations governing the nuclear dynamics:
\beqs
\label{eq:tdhfeq}
i\frac{\partial}{\partial t}\phi_m(\bfx,t) &=&
 -\frac{1}{2} \Delta \phi_m(\bfx,t) - \sum_{(i,j)} \frac{z_j}{|\bfx-\bar{\bfx}_{i,j}(t)|} \phi_m(\bfx,t)\\ & &\quad \quad + \sum_{k=1}^P \bigg(\int_{\R^3} \frac{(\phi^*_{k}\phi_{k})(\bfx^\prime,t)}{|\bfx-\bfx^\prime|}d\bfx^\prime\phi_m(\bfx,t) \nonumber \\
  & & \hspace{3cm} - \int_{\R^3} \frac{(\phi_k^*\phi_m)(\bfx^\prime,t)}{|\bfx-\bfx^\prime|}d\bfx^\prime\phi_k(\bfx,t) \bigg), \nonumber\\
  \label{eq:hfhelfy}
\bff_{i,j} = m_j\frac{d^2 \bar{\bfx}_{i,j}}{dt^2} &=& -\nabla_{\bar{x}_{i,j}}W(\bar{\bfx}_{1,1}(t),\dots, \bar{\bfx}_{1,M}(t), \dots, \bar{\bfx}_{N,M}(t),t ), 
\label{HFmd}
\eeqs
with initial conditions on the single particle wavefunctions, nuclei positions and velocities,
\beq
\phi_m(\bfx,0) = \phi_m^0(\bfx), \;\; \bar{\bfx}_{i,j}(0) = \bar{\bfx}^0_{i,j}, \;\; \frac{d}{dt}\bar{\bfx}_{i,j} (0) =  \bfv_{i,j}^0 ,
\eeq
and $W$ is the potential energy surface expressed in terms of the Slater determinant: 
\beqs
\label{eq:WHF}
W(\bar{\bfx}_{1,1}(t),\dots, \bar{\bfx}_{N,M}(t),t ) = & & -\sum_{(i,j)}\sum_{m=1}^P\int_{\R^3}z_j\frac{(\phi^*_m\phi_m)(\bfx,t)}{|\bfx-\bar{\bfx}_{i,j}|}d\bfx  \\
 & & \hspace{1.5cm} + \frac{1}{2} \sum_{(i,k) \ne(j,l)} \frac{z_k z_l}{|\bar{\bfx}_{i,k} - \bar{\bfx}_{j,l}|}.\nonumber
\eeqs
The dynamical equations for the simulated nuclei under the HF approximation are similar to \eqref{eq:odes_for_ions}, where the electronic density is given by \eqref{eleql0}. The well-posedness and  global-in-time uniqueness of solutions to the above Cauchy problem, for the both the fixed-nuclei and nuclear-dynamics settings, were established by Chadman and Glassey \cite{chadam1975global}, and by Canc\`es and Le Bris \cite{cances1999time}, respectively. In the next section, we prove that these equations are invariant under the action of the affine time-dependent group
in a similar fashion as the general case. 
\subsection{Transformation laws for TDHF equations} \label{sec:invtdhf}
Now we investigate the invariance of Hartree-Fock equation under the group transformation. The group action on the single electron wavefunctions can be derived from the  many-body wavefunction \eqref{gtpsi} using the properties of  Slater determinants. Explicitly,  
\beq
\label{sdgaltrans}
\psi_{\ell}(\bfx_1,  \bfx_2,  \dots, \bfx_P, t) =
e^{-i \bfa_{\ell} \cdot (\bfQ_{\ell}(\bfx_1 + \cdots + \bfx_P) + \frac{1}{2}P \bfa_{\ell} t + \frac{1}{2}P \bfb_{\ell}) }  
 \psi(g_{\ell}^{(t)}(\bfx_1),  g_{\ell}^{(t)}(\bfx_2),  \dots, g_{\ell}^{(t)}(\bfx_P), t) 
 \eeq
 \beqs
 \Longrightarrow \frac{1}{\sqrt{P!}}\begin{vmatrix} \phi^{\ell}_{1}(\bfx_1,t)& \cdots & \phi^{\ell}_{1}(\bfx_{P},t) \\
\vdots & \ddots & \vdots\\
\phi^{\ell}_{P}(\bfx_1,t) & \cdots &\phi^{\ell}_{P}(\bfx_{P},t)\end{vmatrix} & & \nonumber\\
& & \hspace{-6.05cm}= \frac{1}{\sqrt{P!}}\begin{vmatrix} e^{-i \bfa_{\ell}\cdot(\bfQ_{\ell}\bfx_1 + \frac{1}{2}\bfa_{\ell}t + \frac{1}{2} \bfb_{\ell})}\phi_{1}( g_{\ell}^{(t)}\bfx_1,t)& \cdots & e^{-i \bfa_{\ell}\cdot(\bfQ_{\ell}x_P + \frac{1}{2}\bfa_{\ell}t + \frac{1}{2} \bfb_{\ell})}\phi_{1}( g_{\ell}^{(t)}\bfx_{P},t) \\
\vdots & \ddots & \vdots\\
e^{-i a_{\ell}\cdot(\bfQ_{\ell}\bfx_1 + \frac{1}{2}\bfa_{\ell}t + \frac{1}{2} \bfb_{\ell})}\phi_{P}( g_{\ell}^{(t)}\bfx_1,t) & \cdots &e^{-i \bfa_{\ell}\cdot(\bfQ_{\ell}x_P + \frac{1}{2}\bfa_{\ell}t + \frac{1}{2} \bfb_{\ell})}\phi_{P}( g_{\ell}^{(t)}\bfx_{P},t)\end{vmatrix}. \nonumber
\eeqs
Using the multilinearity of the determinant in its columns, the transformation law of each one-electron wave functions $\phi^{\ell}_i$
given
\beq
\phi_i^{\ell}(\bfx_j,t) = e^{-i \bfa_\ell\cdot(\bfQ_{\ell}\bfx_j + \frac{1}{2}\bfa_{\ell}t +  \frac{1}{2} \bfb_{\ell})}\phi_i(g^{(t)}_{\ell}\bfx_j,t)
\label{eq:hfwgt}
\eeq
implies the desired transformation law of the
many-body wave function $\psi_{\ell}$
given in \ref{sdgaltrans}. Therefore, the symmetry-reduced electron dynamics on the Slater manifold $\mathcal{A}$ can be described entirely at the level of the single-electron wavefunctions $\{\phi_m\}_{m=1}^P$. This motivates the following proposition.

\begin{proposition}[Invariance of TDHF under time-dependent group]
\label{pro:pro_tdhfg}
    Suppose that one-electron wavefunction function $\phi_m(\bfx,t) \in \calA$ (see
    (\ref{slatterdet1})) satisfies the time-dependent Hartree Fock equations for all $m \in \{1,\dots,P\}$,
    \beqs
    i\frac{\partial}{\partial t} \phi_m(\bfx,t) &=& \bar{H}_{\psi}^{\bar{\bfx}_{1,1}(t),\dots \bar{\bfx}_{N,M}(t)} \phi_m(\bfx,t),\ \ {\rm where} \nonumber \\
    \ \ \ \bar{H}_\psi^{\bar{\bfx}_{1,1}(t),\dots \bar{\bfx}_{N,M}(t)} &=& -\frac{1}{2}\Delta - \sum_{(i,j)}\frac{z_j}{|\cdot - \bar{\bfx}_{i,j}(t)|} + \left(\sum_{k=1}^P|\phi_k|^2\star\frac{1}{|\bfx|}\right) - \sum_{k=1}^P \left(\phi_k^* \star\frac{1}{|\bfx|}\right)\phi_k 
    \label{HFequation}
    \eeqs
    and with nuclear motions determined by the simulated nuclei:
\beq
\bar{\bfx}_{i,j}(t) = g_i^{(t)}(\bar{\bfx}_{1,j}(t)), \quad i = 1, \dots, N, \quad j = 1, \dots,M, \quad t > 0. 
\label{eq:nuceli_pos}
\eeq
Here, as above, $\calG^{(t)} = \{ g_1^{(t)}, g_2^{(t)}, \dots, g_N^{(t)} \}$ is a discrete group of isometries with affine time dependence of the translations.  Fix $1 \le \ell \le N$, where 
$g_{\ell}^{(t)}(\bfx)= \bfQ_{\ell} \bfx + \bfa_{\ell} t + \bfb_{\ell} \in \calG^{(t)} $, $\bfQ \in$ O(3), $\bfa_{\ell}, \bfb_{\ell} \in \R^3$, and define the group transform of a solution by
\beq
\phi^\ell_m(\bfx,t) = e^{-i\bfa_\ell\cdot(\bfQ_{\ell}\bfx + \frac{1}{2}\bfa_\ell t + \frac{1}{2} \bfb_{\ell})} \phi_m(g_{\ell}^{(t)}\bfx,t), \quad \forall\,  m \in \{1,\dots,P\}.
\label{eq:singleinvar}
\eeq
Then $\phi_m^{\ell}(\bfx,t)$ satisfies \eqref{HFequation}.
\end{proposition}
\begin{proof}
    The proof is similar to the proof of Proposition \ref{ell} and is given in Appendix \ref{sec:ap1} for completeness.
\end{proof}
\begin{lemma}[Uniqueness of single electron wavefunction]
    Suppose $\phi_i(\bfx,t) \in H^1(\R^3\times \R^{\geq0})$   satisfies the time-dependent Hartree-Fock equation,
    \beq
    i\frac{\partial}{\partial t} \phi_i (\bfx,t) = \bar{H}_{\psi}^{\bar{\bfx}_{1,1}(t),\dots \bar{\bfx}_{N,M}(t)}\phi_i(\bfx,t),
    \eeq
     $i\in \{1, \dots, P\}$, subject to the initial condition satisfying the identity 
    \beq
    \phi_i(\bfx,0) = e^{-i\bfa_\ell\cdot\bf(Q_\ell \bfx + \frac{1}{2} \bfb_{\ell})}\phi_i(g_\ell^{(0)}\bfx,0), \quad \forall i \in \{1,\dots, P\}, \quad 1\leq \ell\leq N.
    \label{eq:identity_notime}
    \eeq
    Then $\phi_i$ satisfies the invariance condition 
    \beq
    \phi_i(\bfx,t) = e^{-i\bfa_\ell \cdot(\bfQ_{\ell}\bfx + \frac{1}{2}\bfa_\ell t + \frac{1}{2} \bfb_{\ell})}\phi_i(g_\ell^{(t)}\bfx ,t),
    \label{eq:wvonet}
    \eeq
    for $i \in \{1, \dots, P\}$, $l \in \{1,\dots,N \}$ and $t>0$. 
    \label{lem2}
\end{lemma}
\begin{proof}
    The proof is similar to Lemma \ref{lm:comp}  given in the Appendix \ref{sec:ap1} for completion 
\end{proof}
The Cauchy problem defined by \eqref{eq:tdhfeq} and \eqref{HFmd}, corresponding to the TDHF coupled with the nuclear dynamics, is solved for the  electronic and nuclear degrees of freedom, subject to  time-dependent periodic boundary conditions applied on the one-particle wavefunctions \eqref{eq:wvonet} and nuclei positions \eqref{eq:nuceli_pos}. 

\begin{remark} The single Slater determinant approximation excludes the electron correlation energy. The 
time-dependent full configuration interaction (TD-FCI) gives the exact solution of the TDSE within the selected orbital space \cite{friesecke2003multiconfiguration}. The TD-FCI is based on a linear combination of all possible excited state Slater determinants, $\psi(t) = \sum_I C_I(t) |\Phi_I(t)\rangle$. Here, $\{|\Phi_I(t)\rangle\}_{I=1}^K$ contains all $P$ electron Slater determinants that can be constructed from the chosen orbital basis, including the reference determinant and all possible excitations, and the $C_I \in \C, I = 1, \dots, K$ are time-dependent expansion coefficients. The equations of TD-FCI are also invariant under the affine time dependent isometric groups. Here we briefly show the invariance. Each single-electron wavefunction transforms as \eqref{eq:hfwgt}, hence the determinants ${|\Phi_I(t)\rangle}$  can be similarly written as \eqref{sdgaltrans}. We denote them as $|\Phi_I^\ell(t)\rangle$, giving $\psi_\ell(t) = \sum_IC_{I}(t)|\Phi_I^\ell(t)\rangle$. Inserting these into the TDSE,
\beq
i\sum_I \dot{C}_I(t)|\Phi_I^\ell(t)\rangle + i\sum_IC_I(t) |\dot{\Phi}_I^\ell(t)\rangle = \sum_I C_I(t)H_e^{\bar{\bfx}_{1,1}(t),\dots \bar{\bfx}_{N,M}(t)} |\Phi_I^\ell(t)\rangle.
\eeq
Projecting this onto $|\Phi_J^\ell(t)\rangle$, we get
\beq
i\dot{C}_J(t)  = \sum_I C_I(t)\left(\langle \Phi_J^\ell(t)|H_e^{\bar{\bfx}_{1,1}(t),\dots \bar{\bfx}_{N,M}(t)} |\Phi_I^\ell(t)\rangle - i\langle \Phi_J^\ell(t)|\dot{\Phi}_I^\ell(t)\rangle \right).
\eeq
Following the similar equations as in \eqref{eq:timederi} and \eqref{eq:kinenergy}, the TD-FCI coefficient equations retain their form under the affine time-dependent isometric transformations.
\end{remark}

\section{Numerical formulation and approximations}
\label{sec:alg} 

Having established the invariant structure of NAQD equations under discrete Euclidean isometries with affine time-dependent translations, we now discuss the several features, numerical strategy based on real space finite difference method and approximations to solve the coupled dynamics of electrons and nuclei. 

\subsection{Fundamental domain}
The equations of OQD are solved for the particles inside the time-dependent fundamental domain. The definition and shape of fundamental domain depends on the choice of the group. The simplest case of the isometry group we choose as an example is the three dimensional translation group $G_T$. $G_T$ is generated by the three elements $t_i = (\bfI|\bfe_i)$, $i =1,2,3$, and is given by
\beq
G_T = \{ t_1^{\nu_1}t_2^{\nu_2}t_3^{\nu_3}: \nu_1,\nu_2,\nu_3 \in \Z \} = \{(\bfI|\nu^1\bfe_1+\nu^2\bfe_2+\nu^3\bfe_3): \nu =(\nu^1,\nu^2,\nu^3)\in \Z^3\},
\eeq
where, $\{\bfe_1,\bfe_2,\bfe_3\}$ are the three linearly independent vectors in $\R^3$. For convenience, we choose the integer triple $\nu$ to index the nuclei $\bar{\bfx}_{\nu,k}(t)$, where the identity element associated to $\nu = (0,0,0)$ corresponds to the simulated nuclei $\bar{\bfx}_{(0,0,0),k}(t) = \bar{\bfx}_k (t); k = 1,\dots,M$. To remain consistent with affine-time dependence defined in \eqref{eq:ctimedep}, we replace each basis vectors $\bfe_i$ by $\bfe_i + t \bf\tilde{\bfe}_i$. Without loss of generality, we take $\tilde{\bfe}_i = \bfA\bfe_i, i = 1, 2, 3$, for some linear transformation $\bfA \in \R^{3 \times 3}$. It can be immediately seen that each element of $G_T$ has the affine time dependence and it can be written as,
\beq
G_T^{(t)} = \{(\bfI|\nu^i(\bfI+ t\bfA)\bfe_i): \nu \in \Z^3, i = 1,2,3\}.
\label{eq:tdgt}
\eeq
Here, we used Einstein summation convention. This formulation enables direct comparisons between theories operating at different spatial and temporal scales, as illustrated in Fig.~\ref{fig:time_scale}. The matrix $\bfA$ determines the deformation applied to macroscopic system and is analogous to the Lagrangian description of motion:
\beq
\bfy = (\bfI +t\bfA)\bfX, \quad \text{for} \; \bfX,\bfy \in \R^3.
\eeq
The deformation gradient is $\bfF(t) = \bfI + t\bfA$. The linear transformation $\bfA$ along with the basis vectors $\bfe_i, i = 1,2,3$ are assignable, but they cannot have implicit dependence on time $t$. The only time dependence is explicit as written in \eqref{eq:tdgt}.  The non-simulated nuclei will follow the trajectories restricted by action of group $G^{(t)}_T$ on simulated nuclei,
\beq
\bar{\bfx}_{\nu,k}(t) = g_{\nu}^{(t)}(\bar{\bfx}_{k}(t)) = \bar{\bfx}_k(t) + \nu^i(\bfI + t\bfA)\bfe_i, \quad t>0, \; g_\nu \in G_T^{(t)}.
\label{eq:pos_g}
\eeq
\begin{remark} 
Depending on the choice of $\bfA$, it may occur that for some $t =T>0$, the matrix $(\bfI+t\bfA)$ becomes singular and ceases to be invertible. In this case, we terminate the simulation before reaching time $T$.
\end{remark} 

There may not be a requirement of the fundamental domain in the proposed method as it depends on simulated electrons and nuclei, but for simplicity we assume that the simulated nuclei $\{\bar{\bfx}_{k}(t)\}_{k=1}^M$ belong to the time-dependent fundamental domain $\mathcal{U}\subset \R^3$,
\beq
\mathcal{U}(t) = \{\lambda^i(\bfI + t \bfA )\bfe_i; \; \lambda^i \in [0,1), i = 1,2,3\},
\eeq
attached to the origin and $\mathcal{U}_0$ is the reference domain at $t=0$.
\subsubsection{Redefinition of nuclei to the fundamental domain} \label{sec:redef}
During the simulation, it can happen that the simulated nuclei diffuse out of $\mathcal{U}(t)$ (which causes no issues) but at the same instant a nuclei of same species enters $\mathcal{U}(t)$ with a different velocity, according to \eqref{eq:pos_g}, explicitly,
\beq
\frac{d}{dt}\bar{\bfx}_{\nu,k} (t) = \frac{d}{dt}\bar{\bfx}_k(t) + \nu^i \bfA\bfe_i.\
\label{eq:non-sim-en}
\eeq
For numerical convenience, one may define the non-simulated nuclei $\bar{\bfx}_{\nu,k}$ which entered into $\mathcal{U}(t)$ as a simulated nuclei. Suppose $\bfe^1 , \bfe^2, \bfe^3$ are the reciprocal vectors that satisfy $\bfe^i\cdot \bfe_j = \delta^i_{j}$. The normals to the faces of fundamental domain can be given by $(\bfI+t\bfA)^{T}\bfB = \pm\bfI$ :
\beq
\pm (\bfI +t\bfA)^{-T}\bfe^1, \quad \pm(\bfI +t\bfA)^{-T}\bfe^2, \quad \pm(\bfI +t\bfA)^{-T}\bfe^3,
\eeq
where, columns of matrix $\bfB$ contains these face normal vectors. Since $\bfe_1,\bfe_2, \bfe_3$ are chosen as the linearly independent vectors, a point $\bar{\bfx}_k$ lies in $\mathcal{U}(t)$, if and only if,
\beq
\label{eq:check_sim}
0\leq \bar{\bfx}_{k}\cdot (\bfI + t\bfA)^{-T}\bfe^i<1.
\eeq
If the above condition is violated by any simulated nucleus $\bar{\bfx}_k$, then the nucleus $k$ has passed out of $\mathcal{U}(t)$ and it should be replaced by its image $(\nu,k)$ that has just entered $\mathcal{U}(t)$ with different velocity defined by \eqref{eq:non-sim-en}. So, when solving coupled electrons and nuclei dynamics, the new position and velocity of nuclei have to be used.
\subsubsection{Invariance of one-particle wavefunction} 
The TDHF wavefunctions \eqref{eq:wvonet} under the group $G_T^{(t)}$ will transform as 
\beq
\label{eq:gtwave}
\phi_\alpha (\bfx + \nu^j(\bfI + t\bfA)
\bfe_j,t) = e^{i\left((A\nu^j\bfe_j) \cdot \bfx + \frac{1}{2}\vert A\nu^j\bfe_j\vert^2 t + \frac{1}{2}(\bfA \nu^j\bfe_j)\cdot (\nu^j\bfe_j)\right)}\phi_\alpha(\bfx,t), \quad \bfx \in \mathcal{U}, \; \alpha \in \{1,\dots, P\}.
\eeq
Here, $P$ is the number of states in the unit cell and $\nu^j \in \{-1,1\}\; \text{for} \;j = 1, 2, 3$ depending on the left or right side image of the fundamental domain. The one particle wavefunctions need to be updated at each time step to respect the invariance of the TDHF. From the deconstructionist prespective, the translation $\nu^j(\bfI + t\bfA)\bfe_j$ can be viewed as original translation $\nu^j\bfe_j$ plus the accumulated displacement $t\bfA\nu^j\bfe_j$ with velocity $\bfA\nu^j \bfe_j$. This velocity generates the Gallilean phase.
The exponential prefactor in \eqref{eq:gtwave} maps the wavefunction onto a new domain after the deformation. 

\subsection{Reformulation of governing equations and approximations} 

The non-local and non-linear nature of the Fock exchange term ($V_{ex}$) makes TDHF exceptionally computationally demanding and neglects electron correlation. There is no exact local expression to replace $V_{ex}$ but it can be approximated by locality in both space and time. We therefore employ adiabatic local density approximation (ALDA), where exchange corelation functional depends only on the instantaneous local electron density, $V^{\text{ALDA}}_{xc}(\rho_\psi(\bfx,t))$.  Because the exact exchange--correlation potential depends on the density history and initial state \cite{maitra2002memory,marques2004time}, ALDA neglects memory and cannot generally describe true double excitations, long-range charge transfer, or strongly nonlocal exchange--correlation effects \cite{supp}.

All-electron Coulomb potentials are singular at the nuclei, while rapidly varying core orbitals require prohibitively fine real-space grids \cite{cohen2013locally,beck2000real,rufus2021fast}. Additionally, the tightly bound core electrons are chemically inactive for the non-adiabatic dynamics studied in this paper. We replace the electron--ion interaction with separable nonlocal pseudopotentials in the Kleinman--Bylander representation \cite{Kleinman82,chelikowsky1994finite}. These pseudopotentials represent the combined effect of each nucleus and its frozen core on the valence electrons, producing smoother pseudo-orbitals and substantially reducing the spatial resolution required \cite{supp}. With these approximations, the TDHF evolution is replaced by real-time time-dependent density functional theory (RT-TDDFT). The corresponding time-dependent Kohn--Sham Hamiltonian is
\beq
H_{\rho_{\psi}}^{\bar{\bfx}_{1,1}(t),\dots \bar{\bfx}_{N,M}(t)} = -\frac{1}{2}\Delta + \varphi(\bfx, t)  + V_{xc}^{\text{ALDA}}(\rho_\psi(\bfx,t)) + V_{\text{NL}}(t),
\label{eq:tddft}
\eeq
where, $\varphi$ is the local electrostatic potential generated by the electronic density and ionic pseudocharge distributions and $V_{\text{NL}}$ is non-local pseudopotential operator. 

The numerical implementation of  \eqref{eq:tddft} in time-dependent domain presents two related challenges. First, the electronic structure problem is physically posed on the Eulerian domain $\mathcal{U}(t) = \bfF(t)\mathcal{U}_0 $. A real space finite difference grid attached to $\mathcal{U}(t)$ has physical nodes $y_i(t)$ whose locations change at every time step. Consequently, the wavefunctions would have to be interpolated repeatedly from $\mathcal{U}(t_n)$ to $\mathcal{U}(t_{n+1})$. Such interpolation is not, in general, a unitary operation. It can therefore change the wavefunction norms, introduce artificial smoothing and phase errors, and compromising the time reversibility of the electronic propagation. These errors would be especially difficult to separate from the physical electronic excitation produced by the deformation. The wavefunction norm, however, remains conserved in the continuous TDDFT on deforming FD but  non-conservation may arise as a part of numerical dicretization error (see theorem \ref{thm:sm_eulerian_norm}). 

Second, the group transformation of the wavefunctions introduces the position-dependent phase factor \eqref{eq:gtwave}. 
Under conventional Bloch conditions, the phase associated with a lattice translation is spatially constant, and the factors associated with inverse translations are complex conjugates. This is not generally guaranteed for the position-dependent affine phase. Simply multiplying the finite-difference discretization points outside the FD (ghost nodes) by the time-dependent phase factor while retaining the  standard finite-difference Laplacian may produce boundary matrix elements that do not satisfy $H_{pq} = H^*_{qp}$ (see proposition \ref{prop:sm_eulerian_ghost} in appendix (\ref{sec:sm_pullback_motivation})). This means, that the resulting discrete kinetic operator may fail to be Hermitian. This loss of Hermiticity is a discretization artifact, not a deficiency of the ansatz used in affine transformation of wavefunction (More details are in appendix \ref{sec:sm_pullback_motivation}). 

To avoid these difficulties, we pull the OQD equations back to the fixed Lagrangian domain $\mathcal{U}_0$.  The wavefunction $\phi_n(\bfy,t)$ in Lagrangian coordinate $\bfX$ can be written as

\beq
\widetilde{\phi}_n(\bfX,t) = J(t)^{1/2} \phi_n(\bfF(t)\bfX,t),
\label{eq:wave_rel}
\eeq
where, $\bfy(\bfX,t) = \bfF(t)\bfX$, $J(t) = \det(\bfF(t)) $ is the Jacobian and factor $J^{1/2}$ incorporates the deformation of FD in fixed reference domain and conserves wavefunction inner products between the current and reference domain. Consequently, the electron densities in two frames are related by
\beq
    \widetilde{\rho}_{\psi}(\bfX,t) = J(t)\rho_{\psi}(\bfF(t)\bfX,t).
    \label{eq:densityrel}
\eeq
The map in  \eqref{eq:wave_rel} is unitary and places every time-dependent wavefunctions in the same reference Hilbert space. The deformation then enters through a kinetic operator, the physical positions of simulated and non-simulated nuclei and transformed local and nonlocal potentials. This formulation eliminates repeated interpolation  and permits an explicitly Hermitian discretization.
Direct differentiation of \eqref{eq:wave_rel} gives the general fixed-reference equation (for complete derivation see appendix \ref{sec:sm_pullback})
\beqs
i\frac{\partial \widetilde{\phi}_n}{\partial t} = \widetilde{H}_{\widetilde{\rho}_{\psi}}(t) \widetilde{\phi}_n =  \Bigg[-\frac{1}{2}\nabla_\bfX \cdot \bfG(t)\nabla_\bfX + i\left(\bfc(\bfX,t)\cdot \nabla_\bfX + \frac{1}{2}\nabla_{\bfX}\cdot \bfc(\bfX,t)\right) + \nonumber\\
\widetilde{\varphi}(\bfX,t) + \widetilde{V}_{xc}^{\text{ALDA}}(\widetilde{\rho}_\psi(\bfX,t)) + \widetilde{V}_{\text{NL}}(t) \Bigg]\widetilde{\phi}_n,
\label{eq:lagtddft}
\eeqs
where, $\bfG(t) = \bfF(t)^{-1}\bfF(t)^{-T}$ is the inverse metric appearing in the reference domain gradient and Laplacian, and $\bfc(\bfX,t) = \bfF(t)^{-1}\bfA\bfX$ is reference-space velocity field. The second term inside parenthesis in \eqref{eq:lagtddft} results from change of volume of FD, and $\widetilde{\varphi}(\bfX,t), \; \widetilde{V}_{xc}^{\text{ALDA}}(\widetilde{\rho}_{\psi}(\bfX,t))$ and $\widetilde{V}_{\text{NL}}(t)$ are the electrostatic potential, exchange-correlation functional and non-local pseudopotential in the reference domain. The metric therefore enters the kinetic and Poisson operators, while local
ionic potentials and nonlocal projectors are evaluated using physical
distances such as $| \bfF(t)\bfX -\bar{\bfx}_j(t)|$.

Thus, the orbitals and density are propagated on one fixed grid, while the
metric, physical ion--grid distances, and pseudopotential projectors change
with the deformation and the physical densities, orbitals, currents, and other observables are reconstructed using
\eqref{eq:densityrel} and
\eqref{eq:wave_rel}.
\subsection{Objective quantum dynamics simulations}

In  the coupled evolution of the electron-nuclei systems under the macroscopic deformation, the nuclei are propagated according to the Newtonian dynamics under the forces generated by the time dependent electronic wavefunction and density. The Lagrangian TDDFT equations are discretized on a uniform real space grid using centered finite-difference method and we employ the velocity-Verlet algorithm to compute the nuclei positions and velocities. At $t=0$, the ground state self-consistent calculations are carried out for a prescribed simulated nuclei  $\{\bar{\bfx}_{1,k}(0)\}_{k = 1}^M$ and electrons $\{\bfx_i\}_{i = 1}^P$  in the fundamental domain. Initially, the electronic subsystem may populate an arbitrary state and subsequently evolve non-adiabatically in coupling with the nuclear degrees of freedom. In the present work, the occupations of the electronic orbitals are assumed to remain fixed. This approximation may become inadequate when transitions between electronic states are significant. In such cases, changes in the orbital populations may be incorporated using surface-hopping framework \cite{desouter1979nonadiabatic}.

  The evolution of wavefunction in Lagrangian RT-TDDFT equations \eqref{eq:lagtddft} can be formally written as,
\beqs
\phi_\alpha(\bfx,\mathcal{T}) &=& U(\mathcal{T},t)\phi_\alpha(\bfx,t)\\
&=& \hat{T}\exp{\left(-i\int_t^{\mathcal{T}}\widetilde{H}_{\widetilde{\rho}_{\psi}}(s)ds\right)}\phi_\alpha(\bfx,t), \nonumber
\eeqs
where, $U(\mathcal{T},t)$ is the unitary time-evolution operator (propagator) and $\hat{T}$ is the time ordering operator \cite{cances1999time,sugino1999density}. The above equation provides the formal way of obtaining the wavefunction at any time $t$ but calculating the implicit time dependence of Hamiltonian in \eqref{eq:lagtddft} through the nuclei motion, deformation and density $\rho_{\psi}(t)$ is challenging. It can be dealt by exploiting the composition property of the propagator, $ U(t,s)U(s,r) = U(t,r), \; r \leq s\leq t$. Using the above property, we can write:
\beq
U(\mathcal{T},t) = \prod_{n = 1}^{N}U(t_{n+1},t_n),
\eeq
where, $t_1 = t$, $t_{N+1} = \mathcal{T}$ and $t_{n+1} - t_n = \Delta t$ denote the time step. This allows us to divide the evaluation of the wavefunctions at $\mathcal{T}$ into $N$ short-time propagation. In the case of small enough time step, we use second-order midpoint approximation 
\beq
U(t+\Delta t,t) = \exp\left(-i\Delta t \widetilde{H}_{\widetilde{\rho}_\psi}^{n+1/2}\right) + \mathcal{O}(\Delta t^3),
\label{eq:td_md}
\eeq
where, $\widetilde{H}_{\widetilde{\rho}_{\psi}} = \widetilde{H}_{\widetilde{\rho}_{\psi}}\left[\{\bar{x}_{i,j}^{n+1/2}\},\bfF_{n+1/2},\widetilde{\rho}^{n+1/2}\right]$ is the midpoint Hamiltonian which depends on evolving nuclear geometry, deformation and electronic density and therefore is evaluated at $t_{n+1/2}$ using self-consistent predictor-corrector iteration \cite{Castro2004,ZhuHerbert2018} coupled to velocity Verlet  algorithm. Precisely, with the electronic predictor $q^{(0)} = \widetilde{\rho}_{\psi}^n$, at  corrector iteration $k$, the trial physical midpoint density is
\beq
\widetilde{\rho}_\psi^{n+1/2,(k)} = \frac{1}{2}\left(\frac{\widetilde{\rho}^n}{J_n} + \frac{q^{(k)}}{J_{n+1}} \right), 
\eeq
and nuclear and deformation gradiant midpoint quantities are
\beq
\bar{\bfx}_{1,j}^{n+1/2} = \frac{\bar{\bfx}_{1,j}^{n} + \bar{\bfx}_{1,j}^{n+1}}{2}, \qquad \bfF_{n+1/2} = \bfF\left(t_n + \frac{\delta t}{2}\right),
\eeq
where, $\bar{\bfx}_{1,j}^{n+1}$ is the nuclei position calculated as for a force $\bff_{1,j}^n$ 
\beq
\bar{\bfx}_{1,j}^{n+1} = \bar{\bfx}_{1,j}^{n} + \bfv_{1,j} \Delta t + \frac{\Delta t^2}{2m_j}\bff_{1,j}^n.   
\eeq
Unwrapped nuclear coordinates are used in the position average so that a periodic boundary crossing
does not introduce an artificial midpoint displacement. Consequently, the Crank-Nicolson equation is solved for each $\widetilde{\phi}_\alpha^{n+1,(k)}$ \cite{castro2004propagators,kanungo2019real,meng2008real},
\beq
\left(\bfI + \frac{i\Delta t}{2}\widetilde{H}_{\widetilde{\rho}_{\psi}}^{n+1/2,(k)}\right)\widetilde{\phi}^{n+1,(k)}_{\alpha} = \left(\bfI - \frac{i\Delta t}{2}\widetilde{H}_{\widetilde{\rho}_{\psi}}^{n+1/2,(k)}\right)\widetilde{\phi}^{n,(k)}_{\alpha}.
\eeq
The propagated wavefunctions gives the density $\widetilde{\rho}_{\psi}^{n+1}$, which is mixed using the Anderson mixing scheme \cite{anderson1965iterative}. The procedure is repeated until density reaches self-consistency. After convergence, the  density, potentials, energies, and nuclear forces are re-computed from the accepted wavefunctions. The velocity of simulated nuclei is then completed as
\beq
    \bfv_{1,j}^{n+1}
    =
    \bfv_{1,j}^n
    +
    \frac{\Delta t}{2m_{j}}
    \left(
        \bff_{1,j}^n+\bff_{1,j}^{n+1}
    \right).
    \label{eq:VV_velocity_main}
\eeq
The non-simulated nuclei are generated from the instantaneous deformed lattice.  If any simulated nuclei  diffuse out of $\mathcal{U}(t_n)$, the non-simulated ions of same species enter $\mathcal{U}(t_n)$ at the same instant and they are redefined as simulated ions and explained in section \ref{sec:redef}. This iteration is essential for preserving the time-reversal symmetry and ensuring the long-time stability of the simulation.  Note that the self-consistent procedure used during the time evolution differs from the ground-state self-consistent calculation. In particular, the Hamiltonian is not diagonalized at each time step, and the propagated wavefunctions are not, in general, eigenstates of the instantaneous Hamiltonian. Details of the finite difference method, velocity-Verlet scheme for both simulated and non-simulated nuclei, mixing scheme, linear-solver tolerances, orthonormality thresholds, and step-rejection criteria are given in the Supplemental Material \cite{supp}.

\section{Application} 
\label{sec:application}
\begin{figure}[!htbp]
    \centering
    \subfloat{\includegraphics[scale =0.5,trim={1cm 4cm 0.6cm 2.5cm},clip]{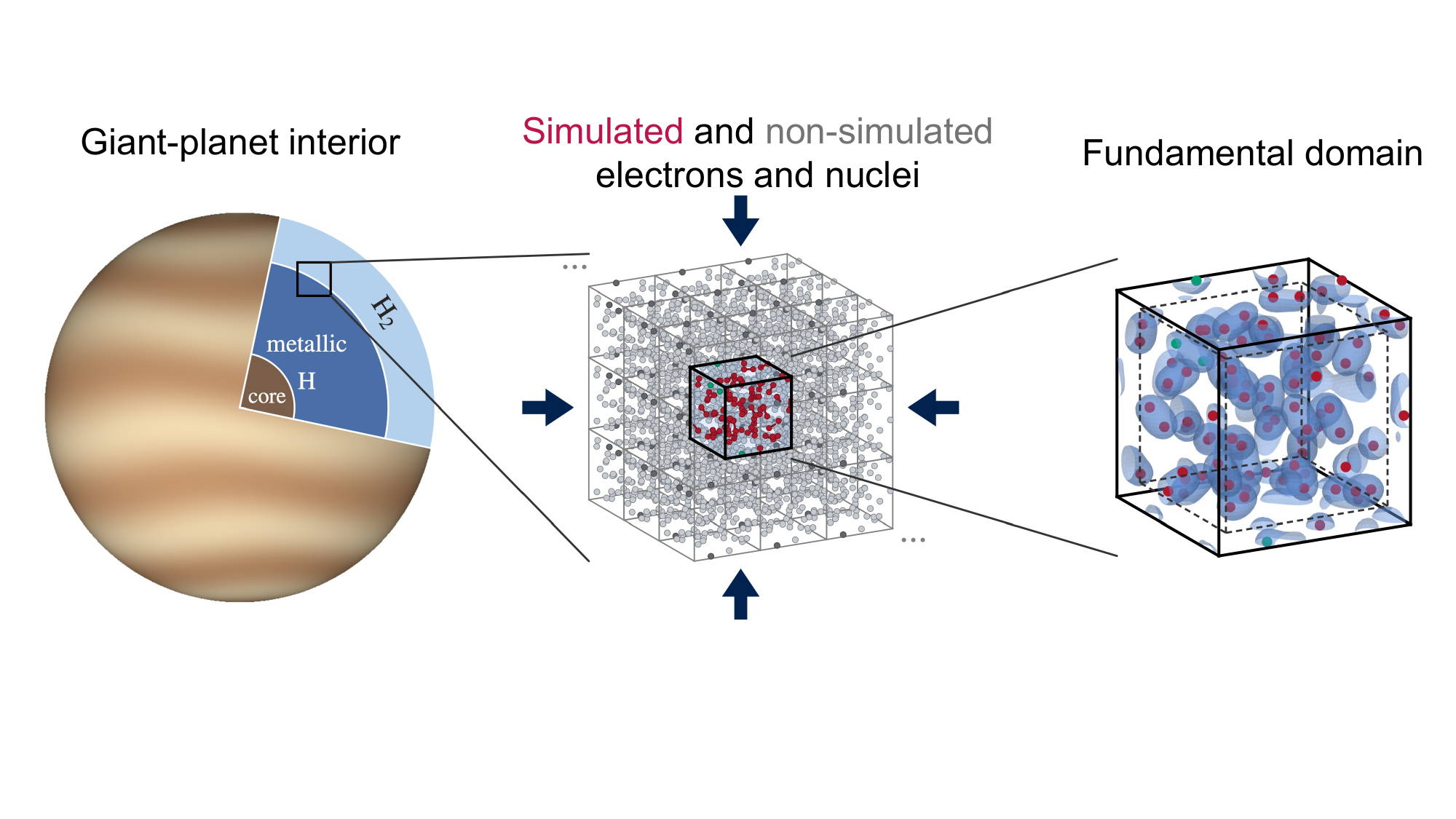}\label{fig:gaint}}
\caption{OQD of hydrogen under rapid compression. Left image is a schematic of giant-planet interior with a molecular H$_2$ envelope above metallic hydrogen. All the electrons and nuclei out to infinity satisfy the equation of non-adiabatic quantum dynamics but only finite set of atoms (shown in red) are simulated. All other atoms (shown in gray) are given by the time-dependent isometry group acting on the simulated atoms. The atoms can diffuse in and out of the fundamental domain.  Arrows are at infinity and indicate the imposed compression.  The right image shows the fundamental domain containing simulated nuclei (red: H$_2$ molecule, green: unpaired H atoms) and electron cloud (blue). The dashed cube marks the compressed domain at some time $t$.}
\label{fig:main_giant}
\end{figure}

We now apply OQD to investigate electron–nuclear dynamics in infinite warm dense hydrogen fluid (Fig.~\ref{fig:main_giant}) subject to sudden compression using time-dependent translation group in \eqref{eq:tdgt}. Hydrogen is the most simple system yet a non-trivial benchmark and of great physical importance. The high-pressure and temperature behavior of Hydrogen determines the evolution of many astrophysical objects including giant gas planet, brown and white dwarf stars and atmosphere of compact stars. \cite{bonitz2024toward,mcmahon2012properties}.  Its high pressure behavior is probed experimentally by dynamic compression \cite{celliers2018insulator}. The first principles simulations of this regime almost always assume that the electrons remain in their instantaneous ground state \cite{morales2010evidence}. Using OQD we can study the non-adiabatic nature of complex electron-nuclei dynamics under macroscopic deformation.

We choose $\bfA= \kappa \bfe_1\otimes\bfe_1 + \kappa \bfe_2\otimes\bfe_2 + \kappa \bfe_3\otimes\bfe_3$, so that $\bfF(t)=(1+\kappa t)\bfI$ describes
isotropic compression. Here, $\kappa$ is the compression rate  and we compare two different compression rates, $\kappa = -4.13 \times 10^{13}$ and $-2.07 \times 10^{13} s^{-1}$
($1\times 10^{-3}$ and $5 \times 10^{-4}$ in atomic units of inverse time). The  fundamental domain is a cube containing 100 simulated hydrogen nuclei and
100 simulated electrons, with initial edge length $6.534$~\AA\ and mass density
$\rho_m = 0.6$~g/cm$^3$), represented in Fig.~\ref{fig:main_giant} at $t=0$. The initial velocity and positions are chosen after equilibrating the system at 1300 K using ab-initio molecular dynamics (AIMD) at 0.6 g/cm$^3$ for 400 fs with a Nosé-Hoover thermostat (time step is 0.1 fs). Over this simulation the pressure remains at 70$\pm 2$ GPa and the ionic temperature is $1300\pm 94$ K (Fig.~S2 in supplementary material \cite{supp}). At the end of AIMD simulation the ionic temperature is 1268K and 48 out of 50 H$_{2}$ molecules are intact (Fig.~\ref{fig:snapshot}(a) \& (d)). This configuration is the starting point of OQD.  The initial electronic state is then prepared by self-consistent ground state DFT calculation of this configuration.  Subsequently,  the electronic states are evolved using  TDDFT in the Lagrangian domain \eqref{eq:lagtddft} with Perdew-Wang parameterization of ALDA \cite{perdew1992accurate}, an optimized norm-conserving Vanderbilt pseudopotential \cite{hamann2013optimized} coupled with velocity-Verlet. We use 12$^{\text{th}}$ central finite difference scheme with mesh spacing of 0.132\AA\ and propagate 70 orbitals (50 orbitals are doubly occupied and 20 orbitals are  initially empty) at the $\Gamma$ point with a time step of $5\times 10^{-4}$fs. Nuclei are propagated with velocity Verlet \eqref{eq:VV_velocity_main} and no thermostat, so all energy deposited in the system comes from the imposed deformation.

\subsection{Ionic response}

\begin{figure}[!htbp]
    \centering
    \subfloat{\includegraphics[scale =0.8,trim={0cm 0cm 0.5cm 0cm},clip]{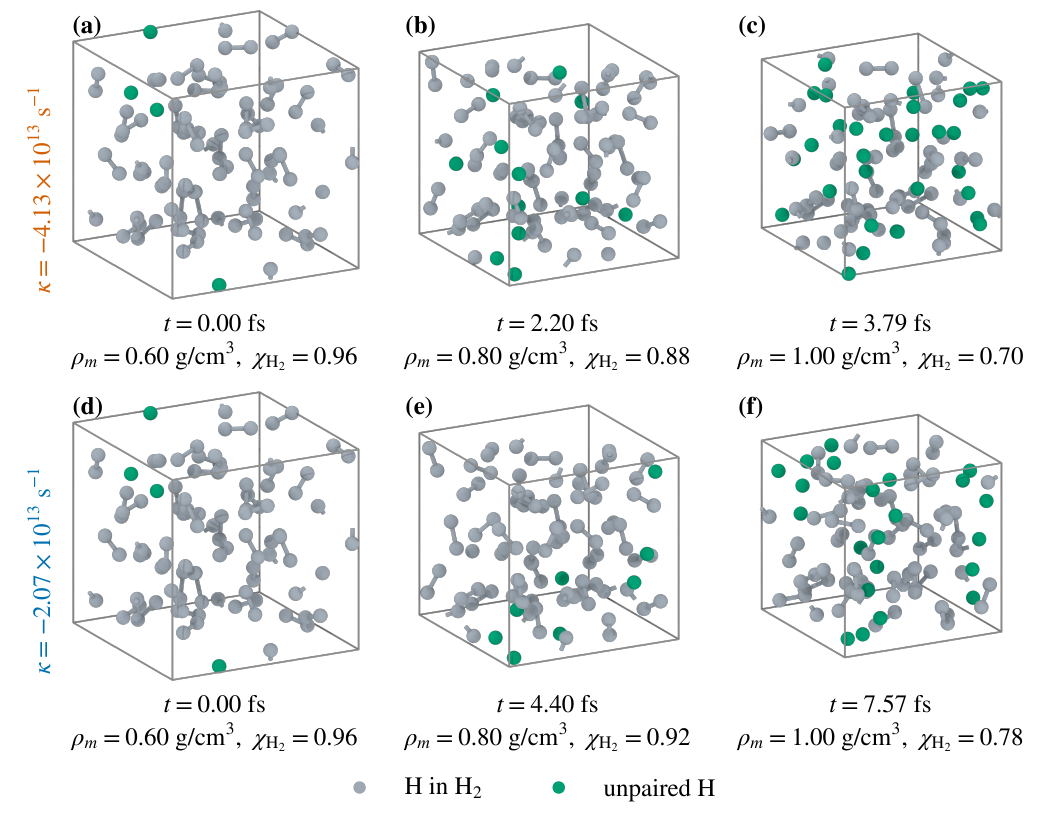}}
\caption{Nuclear configuration during isotropic compression at (a)-(c) $\kappa = -4.13 \times 10^{13}$ and (d)-(f) $\kappa = 2.07 \times 10^{13} $. Both starts from the same initial configuration, so (a) and (d) are identical. Gray atoms belong to H$_2$ molecules (nearest neighbor distance with 0.95 \AA; bonds drawn across periodic boundaries by minimum-image convention), and green atoms are unpaired. Lighter shading indicates greater depth. $\chi_{H_2}$ is the fraction of atoms bound in molecules}
\label{fig:snapshot}
\end{figure}

\begin{figure}[!htbp]
    \centering
         \subfloat[]{\includegraphics[scale =0.65,trim={0.2cm 0cm 0.cm 0cm},clip]{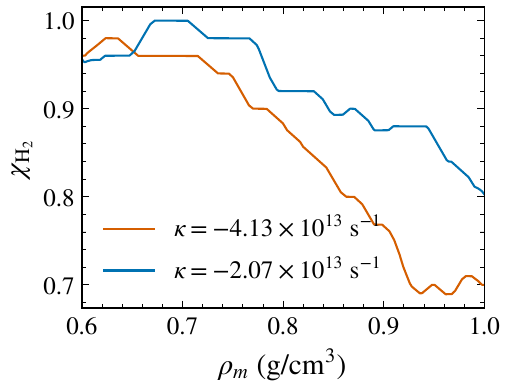}\label{fig:chi}}
       \subfloat[]{\includegraphics[scale =0.65,trim={0.2cm 0cm 0.cm 0cm},clip]{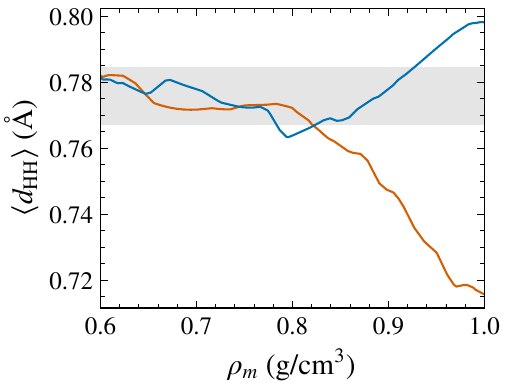}\label{fig:subbl}}
    \subfloat[]{\includegraphics[scale =0.65,trim={0cm 0cm 0cm 0.cm},clip]{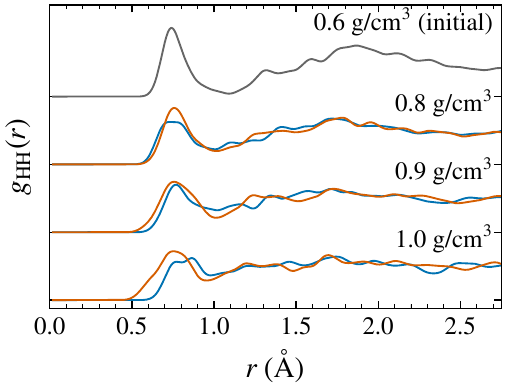}\label{fig:subg}}
\caption{Ionic response at two different compression rates. (a) Fraction of atoms bound in H$_2$ molecule ($\chi_{H_2}$). (b) Mean H–H bond length of the molecules; the gray band is the equilibrium mean of bond length at 1300 K and $\rho_m = 0.6$ g/cm$^3$. (c) H–H pair correlation function at mass density $\rho_m = $ 0.6 (initial structure, gray),  0.8, 0.9, and 1.0 g/cm$^3$ for the two rates.}
\label{fig:waterfall}
\end{figure}

The two compression rate produce distinct nuclear structure. We identify  H$_2$ molecule as a pair of mutual nearest neighbors separated by less than 0.95\AA\ and define  $\chi_{H_2}$ as the fraction of atoms assigned to such pairs. Fig.~\ref{fig:snapshot} shows representative nuclear configurations at intermediate densities for each compression rate, with  unpaired atoms shown in green. During the fast compression, $\chi_{H_2}$ falls below 0.9 at $\rho_m \approx 0.8\mathrm{g/cm^3}$  and reaching approximately 0.70 at  $\rho_m = 1.00\mathrm{g/cm^3}$, compared with 0.78 at the same density during slow compression (Fig.~\ref{fig:chi}). 


\begin{figure}[!htbp]
    \centering
      \subfloat[]{\includegraphics[scale =0.65,trim={0.cm 0cm 0.cm 0cm},clip]{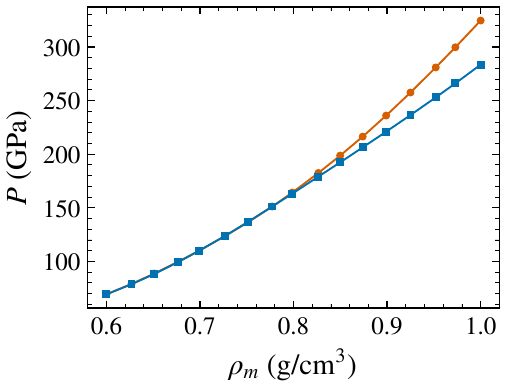}\label{fig:subp}}
    \subfloat[]{\includegraphics[scale =0.65,trim={0.cm 0cm 0.cm 0cm},clip]{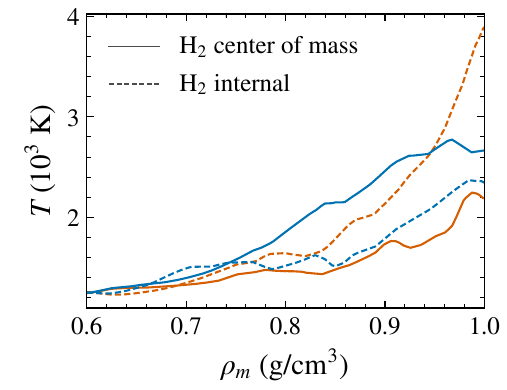}\label{fig:subt}}
\caption{ (a) Instantaneous pressure: Born–Oppenheimer (ground-state) electronic pressure at the $\Gamma$ point plus the
ionic kinetic pressure due the  nuclear velocities. (b) Kinetic temperatures of molecular center-of-mass motion (solid) and internal motion (dashed), from the nuclear velocities with the fitted isotropic flow is removed. }
\label{fig:ionic}
\end{figure}

Fig.~\ref{fig:subg} shows the H--H pair correlation function as the system is compressed at two rates. Initially, the intramolecular peak near 0.74 \AA\ is separated from the intermolecular peak near 1.85 \AA\.   Under compression, the molecules dissociate and atoms come between molecules, leading to flattened intermolecular peak. This is reminiscent of the fluid transition known in these system \cite{su2007excited}. This can also observed by measuring the coordination number. The number of neighbors within $0.95$~\AA{} rises from $0.96$ initially to $1.72$ and $1.44$ at the ends of the fast and slow trajectories, respectively. The fraction of atoms with at least two neighbors within this cutoff increases from zero to $54\%$ and $39\%$, respectively.  At the final density, the intramolecular peak of the fast trajectory develops a tail down to 0.5–0.65 \AA, i.e., bonds compressed below their equilibrium length, whereas that of the slow trajectory is shifted outward with a shoulder at 0.8 \AA. This behavior is consistent with the instantaneous mean bond length measurements, shown in Fig.~\ref{fig:subbl}. It decreases drastically from 0.78 \AA\ to 0.71 \AA\ during the fast compression but for the slow compression the mean bond length moves slightly above the equilibration mean. The abrupt decrease in the mean bond distance during fast compression is accompanied by impulsive excitation of intramolecular motion (vibrational and rotational), where $T_{\text{int}}$ reaches 4000 K in comparison to 2300 K in the slow compression (Fig.~\ref{fig:subt}). While the the center of mass temperatures are $T_{\text{com}} = 2200$ K and 2600 for fast and slow trajectories, respectively. 

Both compression runs are shorter than one H$2$ vibrational period of 8.7 fs (see Fig.~S3 (b) in \cite{supp}). Therefore $\chi_{\text{H}_2}$ measures an instantaneous geometric pairing rather than molecular persistence. Distinguishing transient depairing from persistent bond dissociation would require longer trajectories and pair-resolved analysis of electronic bonding. Such analysis is the scope of future work.

Figure~\ref{fig:subp} shows the pressure estimate $P_{\mathrm{BO}}+P_{\mathrm{ion}}^{\mathrm{kin}}$, obtained by adding the ionic kinetic contribution to the instantaneous BO  electronic pressure evaluated on the OQD configurations. The two pressure curves remain close up to $\rho_m =$ 0.8 g/cm$^3$ and diverge at higher densities, reaching 325 and 284 GPa at $\rho_m =$ 1.0 g/cm$^3$ for fast and slow compression, respectively.   The corresponding ionic kinetic contributions are 28 and 24 GPa and  remaining  difference of 37 GPA arises from the BO electronic contribution evaluated on the different nuclear configurations. This estimate does not include the direct pressure contribution of the nonadiabatic electronic excitations and we will address this in future work.

\subsection{Electronic response}

 We quantify the departure of the system from the instantaneous BO ground state under compression using two measures. First is the number of electrons outside the instantaneous occupied BO  subspace,
\beq
N_{\text{exc}}(t) = N- \sum_{m=1}^{N_o}G_{mm}(t),
\eeq
where, $G_{nm}$ is one-particle density matrix in the instantaneous BO basis:
\beq
G_{mn}(t)= 2 \sum_{j=1}^N\langle \phi_m^{BO}(t)| \phi_j(t)\rangle \langle \phi_j(t)|\phi_n^{BO}(t)\rangle.
\label{eq:denmat}
\eeq
Here, $N$ is the total number of electrons, $N_o =N/2$ is the number of occupied orbitals, and $\{\phi_m^{BO}\}_{m=1}^{70}$ are the BO ground state wavefunctions evaluated at the same instantaneous nuclear configuration. Second, the excess energy, $E_{\text{exc}}= E_{TD}-E_{BO}$, of the propagated state above the ground state. Both quantities vanish in the adiabatic limit. To evaluate these quantities along each compression trajectory, we performed  BO ground state calculations with \(\Gamma\)-point sampling at the 17 nuclear configurations, spanning mass densities from 0.6 to 1 g/cm$^3$ in the step size of 0.025 g/cm$^3$. 

\begin{figure}[!htbp]
    \centering
      \subfloat[]{\includegraphics[scale =0.65,trim={0.cm 0cm 0.cm 0cm},clip]{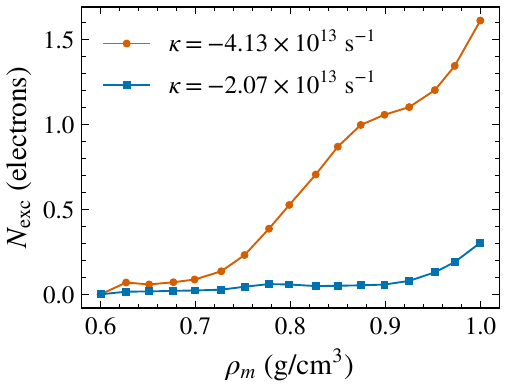}\label{fig:subexcel}}
    \subfloat[]{\includegraphics[scale =0.65,trim={0.cm 0cm 0.cm 0cm},clip]{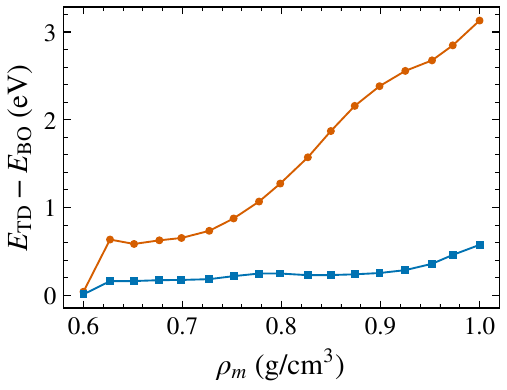}\label{fig:subexcen}}
        \subfloat[]{\includegraphics[scale =0.65,trim={0.cm 0cm 0.cm 0cm},clip]{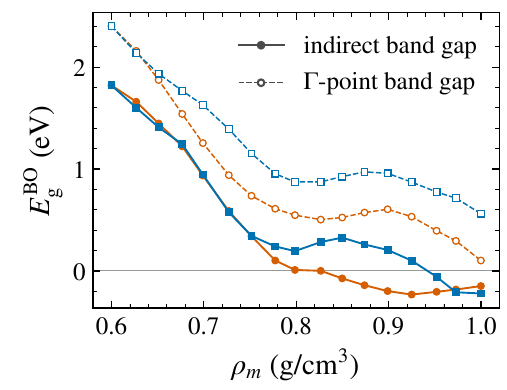}\label{fig:subbandgap}}
\caption{ Electronic response along the compression, evaluated at 17 configurations per rate. (a) Number of electrons outside the instantaneous ground state occupied subspace.  (b) Energy of the propagated state above the  ground state at the same nuclear positions. (c)  Born-Oppenheimer band gap of the same configurations: indirect gap from a $3 \times 3 \times 3$ k-point mesh (filled symbols, solid lines) and gap at the $\Gamma$ point (open symbols, dashed lines).}
\label{fig:electronic}
\end{figure}

\begin{figure}[!htbp]
    \centering
      \subfloat{\includegraphics[scale =0.8,trim={0.cm 0cm 0.cm 0cm},clip]{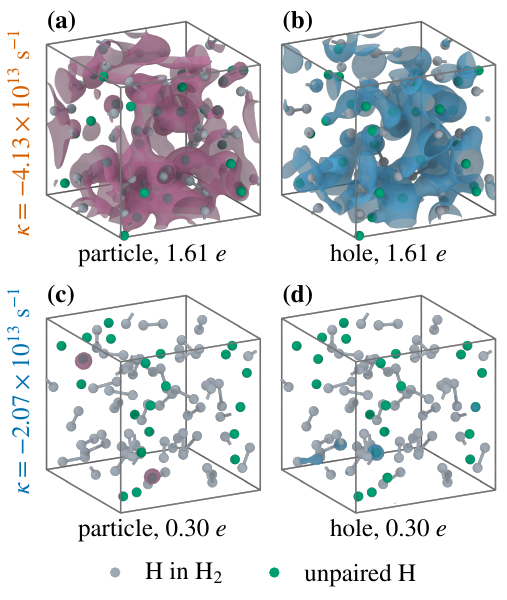}}
\caption{ Particle (a), (c) and hole (b), (d) densities of the
OQD state relative to the instantaneous  Born-Oppenheimer ground
state at 1.0 g/cm$^3$ for (a), (b) $\kappa = 4.13\times10^{13} s^{-1}$ and (c), (d) $\kappa = 2.07 \times 10^{13} s^{-1}$.  }
\label{fig:electronicdensity}
\end{figure}

Figure~\ref{fig:electronic} shows that doubling the compression rate leads to earlier and substantially stronger electronic excitation. At the slower rate, the electronic state remains close to the instantaneous BO ground state up to  $\rho_m \approx 0.90$ g/cm$^{3}$, where $N_{\text{exc}} \approx 0.05$ and $E_{\text{exc}} \approx 0.2$ eV. Beyond this density, both measures slightly increase,  reaching $N_{\text{exc}} = 0.30 $ and $E_{\text{exc}} = 0.57$ eV at the final density of $\rho_m = 1$ g/cm$^3$. At twice the compression rate, the departure from  the BO ground state is already discernible at  $\rho_m =$ 0.72 g/cm$^3$. At the same final density,  $N_{\text{exc}}$ reaches 1.61 and $E_{\text{exc}}$ becomes 3.13 eV, an increase of more than fivefold in comparison to the slower rate (Figs.~\ref{fig:subexcel} and \ref{fig:subexcen}).  To examine the spatial character of the excitation, we decompose the OQD density matrix \eqref{eq:denmat} in the instantaneous BO basis into a particle density carried by states $m>N_o$ and a hole density associated with the states $m\leq N_o$. Fig.~\ref{fig:electronicdensity} shows both densities at $\rho_m= 1.0\ \mathrm{g/cm^{3}}$ using a common absolute isovalue. At this threshold, the fast trajectory exhibits density distributions throughout the fundamental domain, whereas slow rate  encloses only 1–2\% of the excited charge. 

 The onset of this excitation correlates with the evolution of the instantaneous BO band gap at the $\Gamma$ point (Fig.~\ref{fig:subbandgap}). The band gaps are nearly identical initially, when $N_{\text{exc}}$ is also similar at both rates, but the band gap subsequently narrows more rapidly along the faster trajectory. Between $\rho_m \approx 0.83$ and $0.90~\mathrm{g/cm^{3}}$, the gap widens by about $0.1$ eV along both trajectories, and the buildup of excitation slows down but the electrons already excited remian excited. Above $\rho_m \approx 0.90~\mathrm{g/cm^{3}}$, the gap narrows again and both excitation measures resume increasing in both the rates.
The greater excitation at faster driving and its association with gap narrowing are reminiscent of  Landau--Zener transition \cite{zener1932non}, where the transition happen faster when the gap is small and the nuclei move fast.

We next ask whether the electron population transferred above the  occupied BO manifold, obtained by projecting the OQD density matrix  onto the instantaneous $\Gamma$-point BO states, can be described by a Fermi–Dirac (FD) distribution. At the final mass density, $\rho_m=1.0\ \mathrm{g/cm^{-3}}$, the fast trajectory yields a population of  1.61 electrons above the occupied BO states, close to 1.77 electrons predicted by the FD reference $T_{\mathrm{ion}}=4210$ K. For the slow trajectory, the corresponding population is 0.30 electrons, compared with 1.26 electrons predicted by FD reference  at \(T_{\mathrm{ion}}=4260\ \mathrm{K}\). Agreement in the integrated excited-state population, however, does not imply a thermal distribution. At the endpoint of faster compression, the integrated population of instantaneous BO states satisfying $\epsilon_{N_u}-\epsilon_F>2$ eV is 0.08 electrons, compared with only 0.001 electrons in the population-matched Fermi–Dirac reference. This substantial enhancement indicates a nonthermal high-energy tail and related to nonadiabatic phenomena.

 The OQD propagation and the ground state calculations used for above analysis were restricted to the $\Gamma$ point. To assess whether the band gap closes away from $\Gamma$ point and there is an insulator to metal transition under high pressure, we performed additional static BO calculations on the sampled nuclear configurations along each trajectory using a $3\times3\times3$ $k$-point mesh. As shown in Fig.~\ref{fig:subbandgap}, the indirect band gap becomes negative along both trajectories, revealing band overlap that $\Gamma$-only sampling misses. The resulting band overlap is consistent with the onset of metallic behavior in the instantaneous BO electronic structure and would not be detected from the $\Gamma$-point gap alone \cite{hinz2020fully, weir1996metallization,scandolo2003liquid}.

\section{Conclusions and outlook}
\label{sec:conclusions}
In this work, we introduce a novel objective non-adiabatic quantum dynamics (OQD) method for infinite quantum systems undergoing prescribed time-dependent deformation. In this regime, the electrons do not remain on the instantaneous Born-Oppenhiemer manifold. We prove the invariance of quantum dynamics equations by deriving a position dependent phase factor for the transformed wavefunction under the affine time-dependent isometry groups. The method can simulate different kinds of flows such as steady and unsteady compressible and incompressible flows including cases with time-dependent vorticity and cases with strong singularities. We only need to prescribe $3\times 3$ $\bfA$ matrix for different deformation modes, and explicitly simulate only a finite subset of nuclei and electrons.  Every electron and nucleus out to infinity satisfies the equations of non-adiabatic quantum dynamics.  The time-dependent Hartree–Fock and density functional theories also exhibit the same invariance property.  

For numerical implementations, we reformulated OQD in Lagrangian coordinates on a fixed reference fundamental domain and used higher order finite-difference discretization scheme for TDDFT and velocity-Verlet algorithm for ionic evolution. The macroscopic deformation enters through the time dependent deformation gradient  and associated transformed differential operators.  This is done to avoid spurious electronic effects introduced due to repeated interpolation of wavefunctions which can violate norm conservation and orthogonality.  The warm dense hydrogen provides a good trade-off between the physical and chemical realism and tractability. As the application of the method, we studied the coupled electron-nuclear dynamics of 100 simulated hydrogen atoms under two dynamic compression rates. Our simulation underscores a strong rate-dependent non-adiabatic response. At the final density, the fast trajectory has higher non-adiabatic electronic excitation (fig.~\ref{fig:subexcel} \& \ref{fig:subexcen} ) and shows earlier metallization, greater molecular dissociation and different high-density pressure and nuclear response. These calculations demonstrate that OQD can resolve coupled electronic and ionic departures from adiabatic behavior in a deforming infinite material. These results remain a proof of concept because the dynamics use $\Gamma$-point sampling, fixed orbital occupations, a single initial configuration, and compression times shorter than one H\(_2\) vibrational period.

 Future extention of the work include deriving the expressions for stress and pressure within the OQD framework and calculate the transient optical response under deforming system. These developments will allow us to assess how nonadiabatic electronic excitations affect both mechanical and optical observables.  In parallel, we are developing an efficient C/C++ implementation which makes use of domain decomposition and parallelization to improve scaling and computational wall time.  The OQD methodology can be use for investing behavior of other material systems like helium, carbon and mixtures, and study structural and electronic phase transitions driven by different modes of deformation. 
 

\begin{acknowledgments}
This work was supported by the Air Force Office of Scientific Research  under the grant number FA9550-24-1-0344.  It also benefited from the support of the MURI project 
FA9550-25-1-0262. SS and RJD would like to thank Vikram Gavini (University of Michigan), Bikash Kanungo (University of Michigan) and Harley T. Johnson (UIUC)  for insightful discussions and suggestions. Some part of the work presented here by SS was made  during his time at the University of Minnesota. This work used Delta at National Center for Supercomputing Applications through allocation PHY260021 from the Advanced Cyber infrastructure Coordination Ecosystem: Services \& Support (ACCESS) program, which is supported by National Science Foundation grants \#2138259, \#2138286, \#2138307, \#2137603, and \#2138296. We used AI coding assistants Codex (OenAI)  to launch and monitor
jobs on the computing clusters, to make our code efficient, and to prepare the figures. For the  discovery and analysis of the method and basic theorems, AI was not used.   

\end{acknowledgments}

\appendix

\section{Proofs for proposition and lemma in section IV A}
\label{sec:ap1}
\begin{proof}of proposition \ref{pro:pro_tdhfg}. 

The proof is similar to the many-body Schr\"odinger equation. Using \eqref{eq:hfwgt}, we get for the left hand side of the TDHF equation
\beq
i\partial_t \phi^{\ell}_m(\bfx_j,t) = e^{-i \bfa_{\ell}\cdot(\bfQ_{\ell}\bfx_j + \frac{1}{2}\bfa_{\ell}t)}\left(\frac{1}{2}|\bfa_\ell|^2 \phi_m + i\bfa_{\ell}\cdot\phi_{m,\bfx_j} + i\phi_{m,t}\right)(g^{(t)}_{\ell}\bfx_j,t).
\label{eq:tdhf_left}
\eeq
For the kinetic energy:
\beq
\frac{1}{2}\phi^{\ell}_{m,\bfx_j} =\frac{1}{2}e^{-i \bfa_{\ell}\cdot(\bfQ_{\ell}\bfx_j + \frac{1}{2}\bfa_{\ell}t)}\left( -i \bfQ_\ell^T\bfa_{\ell}\phi_m + \bfQ^T_{\ell}\phi_{m,\bfx_j}\right)(g^{(t)}_{\ell}\bfx_j,t).
\eeq
 Differentiating above equation again, we get the Hartree-Fock kinetic energy,
 \beqs
 \frac{1}{2} \phi^{\ell}_{m,\bfx_j\bfx_j} &=&  \frac{1}{2}e^{-i \bfa_{\ell}\cdot(\bfQ_{\ell}\bfx_j + \frac{1}{2}\bfa_{\ell}t)}\left(-|\bfa_{\ell}|^2\phi_m - 2i\bfQ^T_{\ell}\bfa_{\ell}\cdot \bfQ^T_{\ell}\phi_{m,\bfx_j} + \text{tr}(\bfQ^T_{\ell} \phi_{m,\bfx_j\bfx_j}\bfQ_{\ell})\right), \nonumber \\
 &=& e^{-i \bfa_{\ell}\cdot(\bfQ_{\ell}\bfx_j + \frac{1}{2}\bfa_{\ell}t)}\left(-\frac{1}{2}|\bfa_{\ell}|^2\phi_m - i\bfa_{\ell}\cdot \phi_{m,\bfx_j} +  \frac{1}{2}\phi_{m,\bfx_j\bfx_j})\right), 
 \label{eq:HFKE}
 \eeqs
 Combining the \eqref{eq:tdhf_left}  and \eqref{eq:HFKE} gives,
 \beqs
 \label{kinetic_time}
i\frac{\partial}{\partial t} \phi^{\ell}_m (\bfx,t)+ \frac{1}{2} \Delta_{\bfx}\phi^{\ell}_{m} (\bfx,t) & & \\
& &\hspace{-2cm}= e^{-i \bfa_{\ell}\cdot(\bfQ_{\ell}\bfx_j + \frac{1}{2}\bfa_{\ell}t)}\left( i\frac{\partial}{ \partial t} \phi_{m}(g_{\ell}^{(t)}\bfx,t) + \frac{1}{2}\Delta_{\bfx}\phi_{m}(g_{\ell}^{(t)}\bfx,t)\right). \nonumber
\eeqs
Now, we consider nuclear-electron interaction and Hartree potential:
\beq
(V\phi_m^{\ell})(\bfx,t)= -\sum_{(i,j)} \frac{z_j\phi^{\ell}_m(\bfx,t)}{|\bfx-\bar{\bfx}_{i,j}(t)|} + \sum_{k=1}^P \int_{\R^3}\frac{(\phi_k^{\ell*}\phi_k^{\ell})(\bfx^\prime,t)}{|\bfx-\bfx^\prime|}d\bfx^\prime\phi_m^{\ell}(\bfx,t)  \\
\label{eq:pothf}
\eeq
\beqs
&=& e^{-i \bfa_{\ell}\cdot(\bfQ_{\ell}\bfx_j + \frac{1}{2}\bfa_{\ell}t)}\Bigg(-\sum_{(i,j)} \frac{z_j\phi_m(g_{\ell}^{(t)}\bfx,t)}{|\bfx-g_{i}^{(t)}\bar{\bfx}_{1,j}(t)|} +\sum_{k=1}^P \int_{\R^3}\frac{(\phi_k^{*}\phi_k)(g_{\ell}^{(t)}\bfx^\prime,t)}{|\bfx-\bfx^\prime|}d\bfx^\prime\phi_m
(g_{\ell}^{(t)}\bfx,t)\Bigg) \nonumber \\
&=& e^{-i \bfa_{\ell}\cdot(\bfQ_{\ell}\bfx_j + \frac{1}{2}\bfa_{\ell}t)}\Bigg(-\sum_{(i,j)} \frac{z_j\phi_m(g_{\ell}^{(t)}\bfx,t)}{|g_{\ell}^{(t)}\bfx-g_{\ell}^{(t)}g_{i}^{(t)}\bar{\bfx}_{1,j}(t)|} + \sum_{k=1}^P \int_{\R^3}\frac{(\phi_k^{*}\phi_k)(g_{\ell}^{(t)}\bfx^\prime,t)}{|g_{\ell}^{(t)}\bfx-g_{\ell}^{(t)}\bfx^\prime|}d\bfx^\prime\phi_m(g_{\ell}^{(t)}\bfx,t)\Bigg) \nonumber 
\eeqs
\beqs
&=& e^{-i \bfa_{\ell}\cdot(\bfQ_{\ell}\bfx_j + \frac{1}{2}\bfa_{\ell}t)}\Bigg(-\sum_{(i^\prime,j)} \frac{z_j\phi_m(g_{\ell}^{(t)}\bfx,t)}{|g_{\ell}^{(t)}\bfx-g_{i^\prime}^{(t)}\bar{\bfx}_{1,j}(t)|} + \sum_{k=1}^P \int_{\R^3}\frac{(\phi_k^{*}\phi_k)(g_{\ell}^{(t)}\bfx^\prime,t)}{|g_{\ell}^{(t)}\bfx-g_{\ell}^{(t)}\bfx^\prime|}d\bfx^\prime\phi_m(g_{\ell}^{(t)}\bfx,t)\Bigg) \nonumber\\
&=& e^{-i \bfa_{\ell}\cdot(\bfQ_{\ell}\bfx_j + \frac{1}{2}\bfa_{\ell}t)} (V\phi_m)(g_{\ell}^{(t)}\bfx,t). \nonumber
\label{potentialterm}
\eeqs
Finally, considering the non-local exchange interaction, $F$:
\beqs
(F\phi_m^{\ell})(\bfx,t) &=& -\sum_{k=1}^P\int_{\R^3 }\frac{(\phi_k^{\ell*}\phi_{m}^{\ell})(\bfx^\prime,t)}{|\bfx-\bfx^\prime|}dx^\prime\phi_{k}^{\ell}(\bfx,t) \\
&=& e^{-i \bfa_\ell \cdot (\bfQ_{\ell}\bfx + \frac{1}{2}\bfa_\ell t)} \Bigg( -\sum_{k=1}^P \int_{\R^3}\frac{(\phi_k^*\phi_m)(g_\ell^{(t)}\bfx^\prime,t)}{|\bfx-\bfx^\prime|}d\bfx^\prime \phi_k(g_{\ell}^{(t)}\bfx,t)\Bigg) \nonumber \\
&=& e^{-i \bfa_\ell \cdot (\bfQ_{\ell}\bfx + \frac{1}{2}\bfa_\ell t)} \Bigg( -\sum_{k=1}^P \int_{\R^3}\frac{(\phi_k^*\phi_m)(g_\ell^{(t)}\bfx^\prime,t)}{|g_{\ell}^{(t)}\bfx-g_{\ell}^{(t)}\bfx^\prime|}dx^\prime \phi_k(g_{\ell}^{(t)}\bfx,t)\Bigg) \nonumber\\
&=& e^{-i \bfa_\ell \cdot (\bfQ_{\ell}\bfx + \frac{1}{2}\bfa_\ell t)}(F\phi_m)(g^{(t)}_{\ell}\bfx,t). \nonumber
\label{fock_ex}
\eeqs
Here, we used closure property of the group, $g_{\ell}^{(t)}\calG^{(t)} = \calG^{(t)}$.
This with \eqref{kinetic_time} and (\eqref{eq:pothf} completes the proof.
\end{proof}

\begin{proof} of lemma \ref{lem2}.

Define $\phi_i$ by  \eqref{eq:singleinvar}. Then  \eqref{eq:identity_notime} implies that
    \beq
    \phi_i(\bfx,0) = \phi^\ell_i(\bfx,0), \quad \forall i \in \{1,\dots, p\}.
    \eeq
    Since $\phi_i$ and $\phi^{\ell}_i$ both satisfy the same TDHF equation with the same initial conditions then by uniqueness, they are same.
\end{proof}

\section{Eulerian formulation and motivation for Lagrangian domain}
\label{sec:sm_pullback_motivation}
The translation group $G_T^{(t)}$ used for the deforming lattice is
\beq
g_{\nu^j}^{(t)}(\bfy)
=
\bfy+\nu^j(\bfI + t\bfA)\bfe_j,
\qquad
\nu^j\in\mathbb{Z},
\label{eq:affine_translation}
\eeq
which gives the macroscopic deformation
\beq
 \bfy=\bf(\bfI + t\bfA)\bfX,
 \eeq
 with deformation gradient and Jacobian
 \beq
 \bfF(t)=\bfI+t\bfA,
 \qquad
 J(t)=\det\bfF(t)>0,
 \label{eq:sm_motion}
\eeq
where $\bfX\in\calU_0$ is a point in reference fundamental domain (FD) and
$\bfy\in\calU(t)=\bfF(t)\calU_0$ is the point in deformed FD.  Define $\bfT^\nu_i = \nu^i\bfF(t)\bfe_i = \nu^i(\bfI+\bfA t)\bfe_i$ and $  \nu = \{\pm 1\}$ for the nearest left and right images and effect of $G_T^{(t)}$ on wavefuntion is
\beq
\phi_\alpha(\bfy + \bfT_i^\nu,t) = e^{\Theta_{\nu,i}(\bfy,t)}\phi_\alpha(\bfy,t),
\label{eq:phsewave}
\eeq
where, $\Theta_{\nu,i}(\bfy,t) = (\bfA\nu^i\bfe_i)\cdot\bfy
   +\frac12\lvert\bfA\nu^i\bfe_i\rvert^2t
   +\frac12(\bfA\nu^i\bfe_i)\cdot\nu^i\bfe_i$ is the position dependent affine phase introduced in the main text. 

At every fixed time $t$, let the Hamiltonian in Eulerian frame,
\beq
H_{\rho_{\psi}}^{\bar{\bfx}_{1,1}(t),\dots \bar{\bfx}_{N,M}(t)} = -\frac{1}{2}\Delta_y + \varphi(\bfy, t)  + V_{xc}^{\text{ALDA}}(\rho_\psi(\bfy,t)) + V_{\text{NL}}(t),
\eeq
acts on suitable functions in $H^1(\R^3;\C)$. Then because $\varphi(\bfy,t)$ and $V_{xc}^{\text{ALDA}}$ are real, and non-local projectors are Hermitian, we get
\beq
\langle v, H_{\rho_\psi}u\rangle - \langle H_{\rho_\psi}v,u\rangle = -\frac{1}{2} \int_{\R^3} \left(v^*\nabla^2 u - (\nabla^2 v)^* u\right) d\bfy.
\label{eq:greenfun}
\eeq
For the kinetic term Green's identity gives
\beq
 -\frac12 \lim_{R\to \infty}\int_{\partial B_R}
 \left(
 v^*\partial_nu-(\partial_nv)^*u
 \right)\text{d} S =0,
 \label{eq:sm_green_boundary_form}
\eeq
where, $B_R$ is a ball of radius $R$. Therefore,
\beq
\langle v, H_{\rho_\psi}u\rangle = \langle H_{\rho_\psi}v,u\rangle .
\eeq
However, for bare  Hamiltonian operator restricted to  the deforming FD $\calU(t)$ with position dependent affine phase, we can not generally prove $H_{\rho_{\psi}} = H_{\rho_{\psi}}^\dagger$. 

\begin{theorem}[Norm conservation in the deforming fundamental domain]
\label{thm:sm_eulerian_norm}
Suppose that \(\phi_\alpha\) satisfies the Eulerian Schr\"odinger equation and
the affine image relation on every pair of faces of \(\calU(t)\).  Assume that
the local potential is real and that any nonlocal term is Hermitian.  Then
\beq
 N_\alpha(t)
 =
 \int_{\calU(t)}
 \left|\phi_\alpha(\bfy,t)\right|^2 d\bfy
\nonumber
\eeq
is independent of time.
\end{theorem}

\begin{proof}
For the kinetic and local-potential terms, the Schr\"odinger equation and its
complex conjugate give the continuity equation
\beq
 \frac{\partial}{\partial t}\left|\phi_\alpha\right|^2
 +
 \nabla_{\bfy}\!\cdot
 \Im\!\left(
 \phi_\alpha^*\nabla_{\bfy}\phi_\alpha
 \right)
 =0.
\nonumber
\eeq
The Hermitian nonlocal pseudopotential terms contribute zero to the change of the
integrated norm.  Since the domain is time dependent then using the Reynolds transport theorem and continuity equation, we get
\beqs
 \frac{d N_\alpha}{d t}
 &=&
 \int_{\calU(t)}
 \frac{\partial}{\partial t}
 \left|\phi_\alpha\right|^2 d\bfy
 +
 \int_{\partial\calU(t)}
 \left|\phi_\alpha\right|^2
 \frac{d\bfy}{d t}\cdot\mathbf n\,d S \nonumber\\
 &=&
 -\int_{\partial\calU(t)}
 \left(
\Im\!\left(
 \phi_\alpha^*\nabla_{\bfy}\phi_\alpha
 \right)
 -
 \left|\phi_\alpha\right|^2\frac{d\bfy}{d t}
 \right)\cdot\mathbf n\,d S. \nonumber
\eeqs
The second term in the first equation above is due to moving boundary. It remains to compare the integrand on paired boundaries.  Considering boundary along $\bfe_1$ then corresponding boundary
points are \(\bfy(t) \in \partial \calU_1^-\) and \(\bfy(t)+\bfT^+_1(t)\in \partial \calU_1^+\), and their velocities
satisfy
\beq
 \frac{d}{d t}\bigl(\bfy(t)+\bfT^+_1(t)\bigr)
 =
 \frac{d\bfy(t)}{d t}+\bfA\bfe_1.
\nonumber
\eeq
The image relation and its gradient from \eqref{eq:diffimag} give, 
\beqs
 &&\left|\phi_\alpha(\bfy+\bfT^+_1,t)\right|^2
 =\left|\phi_\alpha(\bfy,t)\right|^2, \nonumber\\
 &&\Im\!\left[
 \phi_\alpha(\bfy+\bfT^+_1,t)^*
 \nabla_{\bfy}\phi_\alpha(\bfy+\bfT^+_1,t)
 \right] \nonumber\\
 &&\quad=
\Im\!\left[
 \phi_\alpha(\bfy,t)^*\nabla_{\bfy}\phi_\alpha(\bfy,t)
 \right]
 +\bfA\bfe_1\left|\phi_\alpha(\bfy,t)\right|^2. \nonumber
\eeqs
Consequently, the probability flow relative to the plus moving boundary is
\beqs
 &&\Im\!\left[
 \phi_\alpha(\bfy+\bfT^+_1,t)^*
 \nabla_{\bfy}\phi_\alpha(\bfy+\bfT^+_1,t)
 \right]
 -\left|\phi_\alpha(\bfy+\bfT^+_1,t)\right|^2
 \frac{d}{d t}\bigl[\bfy(t)+\bfT^+_1(t)\bigr] \nonumber\\
 &&\quad=
\Im\!\left[
 \phi_\alpha(\bfy,t)^*\nabla_{\bfy}\phi_\alpha(\bfy,t)
 \right]
 -\left|\phi_\alpha(\bfy,t)\right|^2
 \frac{d\bfy(t)}{d t}. \nonumber
\eeqs
Specifically, the left-hand side is the probability-flow vector relative to \(\bfy+\bfT^+_1 \in \partial\calU_1^+\), and the right-hand side is the
probability-flow vector relative to \(\bfy \in \partial\calU_1^-\).  The equality
shows that these two relative-flow vectors are identical.  If \(\mathbf n\)
is the outward normal of the $\partial\calU_1^+$ , then \(-\mathbf n\) is the outward
normal of the $\partial\calU_1^-$, and their combined contribution is
\beq
 \bigl[\text{relative flow at $\partial\calU_1^+$}\bigr]\cdot\mathbf n
 +
 \bigl[\text{relative flow at $\partial\calU_1^-$}\bigr]\cdot(-\mathbf n)
 =0.
\nonumber
\eeq
Repeating the same argument for all three boundary pairs $ \partial \calU_\nu^{\pm}$ gives
\(d N_\alpha/d t=0\).
\end{proof}

Theorem~\ref{thm:sm_eulerian_norm}  establishes conservation of the wavefunction norm for the complete evolution on the moving fundamental domain. This result should not be confused with Hermiticity of the bare Eulerian Hamiltonian restricted to a frozen instantaneous FD.  On a fixed domain $\calU$,
\(N(t)=\int_{\mathcal U}|\phi|^2\,d\mathbf y\)
has derivative 
has derivative
\[
\frac{dN}{dt}
=
i\langle\phi,(H^\dagger_{\rho_\psi}-H_{\rho_{\psi}})\phi\rangle.
\]
Therefore, on a fixed domain, norm conservation is determined solely by  \(H_{\rho_\psi}^\dagger=H_{\rho_\psi}\). On the deforming domain \(\calU(t)\), however, differentiation of the norm also produces the Reynolds boundary-motion term. The wavefunction relation \eqref{eq:phsewave} makes this term cancel through each pair of boundary components. Consequently, the complete deforming domain evolution conserves probability even though the bare instantaneous Laplacian on FD does not satisfy the ordinary fixed-domain Hermiticity condition.

We now examine whether the  Hamiltonian restricted to one instantaneous FD is Hermitian. The direct differentiation of $\eqref{eq:phsewave}$ gives
\beq
 \nabla_{\bfy}\phi_\alpha(\bfy+\mathbf{T},t)
 =
 e^{i\Theta_{+1}(\bfy,t)}
 \left(
  \nabla_{\bfy}\phi_\alpha(\bfy,t)
  +i\bfA\bfe_1\phi_\alpha(\bfy,t)
 \right).
\label{eq:diffimag}
\eeq
Let the normal in \(\partial_n\) point from the FD boundary containing \(\bfy\)
toward the paired image boundary containing \(\bfy+\mathbf{T}\).  Substitution for two
wavefunctions in \eqref{eq:greenfun} gives
\beqs
 &&\left(
 \phi_\beta^*\partial_n\phi_\alpha
 -(\partial_n\phi_\beta)^*\phi_\alpha
 \right)_{\bfy+\bfT^+_1 \in \partial\calU_1^+}
 \nonumber\\
 &&\quad=
 \left(
 \phi_\beta^*\partial_n\phi_\alpha
 -(\partial_n\phi_\beta)^*\phi_\alpha
 \right)_{\bfy \in \partial\calU_1^-}
 +2i(\bfA\bfe_1\cdot\mathbf n)
 \phi_\beta(\bfy,t)^*\phi_\alpha(\bfy,t). \nonumber
\eeqs
The two boundaries have opposite outward normals, so the first terms on both sides cancel but
the final term on right hand side remains.  Thus the bare Green boundary form does not generally
vanish; it does so only in the special case
\(\bfA\bfe_1\cdot\mathbf n=0\).

\begin{proposition}[Hermiticity of the direct finite-difference scheme in Eulerian frame]
\label{prop:sm_eulerian_ghost}
A centered finite-difference Laplacian constructed by applying the affine
phase only to ghost nodes is not generally Hermitian at finite grid spacing.
\end{proposition}

\begin{proof}
Let a periodic finite-difference grid contain the $N$ points
\(\bfy_0,\ldots,\bfy_{N-1}\), with
$
\bfy_m
=
\bfy_0+m\bfh,
\;
m=0,\ldots,N-1,$ where $\bfh = \frac{\bfT(t)}{N}$ is the grid step and
\beq
 \bfy_{N-1}-\bfy_0=\mathbf{T}(t)-\mathbf{h} .
 \label{eq:sm_last_grid_point}
\eeq
The affine phase is applied to the two ghost nodes of, for example, second-order central finite-difference discretized Laplacian, $L_{E} = \Delta$, to map it back to fundamental domain then   the two corner entries have the form
\beqs
 (L_{\mathrm E})_{0,N-1}
 &=&c_h e^{i\Theta_{-1}(\bfy_{N-1},t)},\\
 (L_{\mathrm E})_{N-1,0}
 &=&c_h e^{i\Theta_{+1}(\bfy_0,t)},
 \label{eq:sm_eulerian_corner_entries}
\eeqs
where \(c_h\) is real and scales as \(h^{-2}\).  Hermiticity requires
\beq
 e^{i\Theta_{-1}(\bfy_{N-1},t)}
 =
 e^{-i\Theta_{+1}(\bfy_0,t)} .
 \label{eq:sm_corner_hermiticity}
\eeq
For the affine phase used here, Eq.~\eqref{eq:sm_last_grid_point} instead
gives
\beq
 \Theta_{-1}(\bfy_{N-1},t)
 +\Theta_{+1}(\bfy_0,t)
 =
 (\bfA\bfe_1)\cdot\mathbf{h}
 \quad (\mathrm{mod}\ 2\pi),
 \label{eq:sm_discrete_phase_mismatch}
\eeq
and hence
\beq
 e^{i\Theta_{-1}(\bfy_{N-1},t)}
 e^{i\Theta_{+1}(\bfy_0,t)}
 =
 e^{i(\bfA\bfe_1)\cdot\mathbf{h}},
 \label{eq:sm_discrete_phase_product}
\eeq
which is not generally unity.  Therefore,
\beq
 L_{\mathrm E} \neq L_{\mathrm E}^{\dagger}.
 \label{eq:sm_discrete_nonhermiticity}
\eeq
\end{proof}

We have used second order central finite difference as the illustration but it will fail for any order of central finite difference. This is the discrete counterpart of the uncanceled instantaneous  boundary term
above: phase-filling the ghost nodes of the bare Laplacian does not include
the motion of the FD boundary. 

For a fine grid,
\beq
 e^{i(\bfA\bfe_1)\cdot\mathbf{h}}-1
 =
 i(\bfA\bfe_1)\cdot\mathbf{h}
 +\mathcal O(|\mathbf{h}|^2).
 \label{eq:sm_phase_mismatch_expansion}
\eeq
Thus the dimensionless phase mismatch, and possibly a Hermiticity residual
normalized by the full operator norm, decreases under refinement.  This does
not restore exact Hermiticity at finite spacing.  Since
\(c_h=\mathcal O(h^{-2})\), the absolute corner-entry difference behaves as
\beq
 \left|
 (L_{\mathrm E})_{0,N-1}
 -(L_{\mathrm E})_{N-1,0}^*
 \right|
 =
 |c_h|\mathcal O(h)
 =
 \mathcal O(h^{-1})
 \label{eq:sm_absolute_corner_mismatch}
\eeq
for a second-order stencil. 

The Hermiticity error above is specific to the direct boundary phase finite-difference construction and is not an intrinsic failure of Schr\"odinger equation in the deforming domain. At any instantaneous time, one can instead construct a Hermitian spatial operator by interpreting the affine phase factor as a $U(1)$ gauge transition function and introducing compatible real-valued gauge connection $\bfa(\bfy,t)$.

\begin{proposition}[Hermiticity of the instantaneous gauge-invariant spatial operator]
\label{prop:sm_covariant_eulerian}
Let a wavefunction obey
\beq
 \phi(\bfy+\mathbf T_\nu,t)
 =e^{i\Theta_\nu(\bfy,t)}\phi(\bfy,t).
 \label{eq:sm_covariant_transition}
\eeq
Suppose there exist a real-valued connection \(\mathbf a(\bfy,t)\) satisfing
\beq
 \bfa(\bfy+\bfT_\nu,t)
 =\bfa(\bfy,t)+\nabla_{\bfy}\Theta_\nu(\bfy,t).
 \label{eq:sm_connection_transition}
\eeq
Then the effective gradient operator
\beq
 [\mathcal D\phi](\bfy,t)
 =
 \left[\nabla_{\bfy}-i\mathbf a(\bfy,t)\right]\phi(\bfy,t)
 \label{eq:sm_covariant_derivative}
\eeq
obeys the same time-dependent transformation relation as the wavefunctions, and the quadratic form associated with the covariant kinetic operator is Hermitian at fixed time t. A discretization of this kintetic operator constructed as $T^h_{\mathcal D} = \frac{1}{2}D_h^\dagger D_h$ is therefore Hermitian at finite grid
spacing, where $D_h$ is the finite-difference matrix.
\end{proposition}

\begin{proof}
Using Eqs.~\eqref{eq:sm_covariant_transition} and
\eqref{eq:sm_connection_transition} gives
\beq
 \begin{split}
 [\mathcal D\phi](\bfy+\mathbf T_\nu,t)
 &=\left[\nabla_{\bfy}-i\mathbf a(\bfy+\mathbf T_\nu,t)\right]
   e^{i\Theta_\nu(\bfy,t)}\phi(\bfy,t)\\
 &=e^{i\Theta_\nu(\bfy,t)}
   \left[\nabla_{\bfy}-i\mathbf a(\bfy,t)\right]\phi(\bfy,t)\\
 &=e^{i\Theta_\nu(\bfy,t)}
   [\mathcal D_{\bfy}\phi](\bfy,t).
 \end{split}
 \label{eq:sm_covariant_derivative_transition}
\eeq
Consequently, \(\mathcal D\phi\) obeys the same affine image relation as \(\phi\) on paired boundaries. 

To show the boundary cancellation explicitly, let \(\mathbf n\) be the
outward normal at \(\bfy+\mathbf T_\nu \in \partial \cal_1^+\)and the outward normal at the paired
point \(\bfy \in \partial \calU_1^-\) is then \(-\mathbf n\).  For the two wavefunctions
\(\phi_\alpha\) and \(\phi_\beta\),
Eq.~\eqref{eq:sm_covariant_derivative_transition} and their affine image
relations give
\beqs
 &&\left(
 \phi_\beta^*\,\mathbf n\!\cdot\mathcal D\phi_\alpha
 -\left(\mathbf n\!\cdot\mathcal D\phi_\beta\right)^*\phi_\alpha
 \right)_{\bfy+\mathbf T_\nu \in \partial \calU_1^+ }
 \nonumber\\
 &&\quad=
 e^{-i\Theta_\nu(\bfy,t)}\phi_\beta(\bfy,t)^*\,
 \mathbf n\!\cdot
 e^{i\Theta_\nu(\bfy,t)}[\mathcal D\phi_\alpha](\bfy,t)
 \nonumber\\
 &&\qquad-
 \left\{
 \mathbf n\!\cdot
 e^{i\Theta_\nu(\bfy,t)}[\mathcal D\phi_\beta](\bfy,t)
 \right\}^*
 e^{i\Theta_\nu(\bfy,t)}\phi_\alpha(\bfy,t)
 \nonumber\\
 &&\quad=
 \left(
 \phi_\beta^*\,\mathbf n\!\cdot\mathcal D\phi_\alpha
 -\left(\mathbf n\!\cdot\mathcal D\phi_\beta\right)^*\phi_\alpha
 \right)_{\bfy \in \partial \calU_1^-}
 \nonumber\\
 &&\quad=-
 \left(
 \phi_\beta^*\,(-\mathbf n)\!\cdot\mathcal D\phi_\alpha
 -\left((-\mathbf n)\!\cdot\mathcal D\phi_\beta\right)^*\phi_\alpha
 \right)_{\bfy \in \partial \calU_1^-}.
 \label{eq:sm_covariant_boundary_cancellation}
\eeqs

Repeating this calculations for every boundary pair makes the complete boundary integral vanish.  The kinetic-energy
matrix elements therefore satisfy
\beq
 \frac12\int_{\calU(t)}
 \left(\mathcal D\phi_\beta\right)^*\!\cdot
 \mathcal D\phi_\alpha\,d\bfy
 =
 \left[
 \frac12\int_{\calU(t)}
 \left(\mathcal D\phi_\alpha\right)^*\!\cdot
 \mathcal D\phi_\beta\,d\bfy
 \right]^*.
 \label{eq:sm_covariant_quadratic_form}
\eeq
This is the Hermiticity condition for the instantaneous kinetic-energy
operator on wavefunctions satisfying the affine time-dependent phase factor relations.

On a grid, associate with every oriented edge \(g\to g'\) the parallel
transport factor
\beq
 U_{gg'}
 =\exp\!\left[-i\int_{\bfy_g}^{\bfy_{g'}}
 \mathbf a(\boldsymbol\ell,t)\cdot d\boldsymbol\ell\right].
 \label{eq:sm_covariant_link}
\eeq
Reversing the edge reverses the line integral, so
\beq
 U_{g'g}=U_{gg'}^*.
 \label{eq:sm_reverse_link}
\eeq

For an edge of length \(h_{gg'}=|\bfy_{g'}-\bfy_g|\), define the
phase-compatible finite difference discretization of gradient operator for the grid values of \(\phi_\alpha\):
\beq
 (D_h\phi_\alpha)_{gg'}
 =
 \frac{U_{gg'}\phi_{\alpha,g'}-\phi_{\alpha,g}}{h_{gg'}}.
 \label{eq:sm_phase_compatible_difference}
\eeq
The kinetic matrix
\(T_{\mathcal D}^h=\frac12(D_h)^\dagger D_h\) then satisfies
\(T_{\mathcal D}^h=(T_{\mathcal D}^h)^\dagger\).
\end{proof}

Mathematically, effective gradient operator has the same structure as the electromagnetic vector potential, although here \(\bf a\) is introduced to represent the affine time-dependent phase geometry and is not necessarily a physical electromagnetic field.

Proposition~\ref{prop:sm_covariant_eulerian} is an instantaneous spatial
construction that is the FD are held fixed at a
given time.  It does not state that the bare Laplacian defined on the deforming FD with
phase-filled ghost values is Hermitian, nor does it by itself establish a
time-propagation scheme dynamically equivalent to the original
Schr\"odinger equation.  Such an numerical scheme on deforming domain would additionally require
the time derivative, time-dependent affine phase factor, boundary motion, and
deforming-domain quadrature to be discretized consistently so that the
exact cancellations between the temporal- and spatial-phase terms, together
with the norm balance, are retained. This is very cumbersome and we do not
implement this alternative.  Instead, all calculations use the
fixed-reference frame  which incorporates these ingredients
in a single fixed-domain representation.

To perform calculations in Lagrangian frame, we need to transform  coordinates, differential
operator, measure, and boundary conditions together. Let $\mathscr{U}(t)$ denote the complete transformation  to the Lagrangian frame, $\widetilde{\phi}_\alpha=\mathscr{U}(t)\phi_\alpha$, the reference Hamiltonain  is
\beq
 \widehat{H}(t)
 =
 \mathscr{U}(t)H(t)
\mathscr{U}(t)^\dagger
 +
 i\dot{\mathscr{U}}(t)
\mathscr{U}(t)^\dagger .
 \label{eq:sm_unitary_generator}
\eeq
The second term is required because the transformation is time dependent. Omitting it
would give an equation that is not equivalent to the Eulerian TDSE.

\subsection{Lagrangian representation}
\label{sec:sm_pullback}

Here, we derive the TDDFT in Lagrangian coordinates. The wavefunctions in Lagrangian frame can be written as
\beq
\widetilde{\phi}_\alpha(\bfX,t) = J(t)^{1/2}\phi_\alpha(\bfF(t)\bfX,t)
\label{eq:wavetrans}
\eeq
The factor $J^{1/2}$ preserves the orbital inner product:
\beq
 \int_{\calU(t)}\phi_\alpha^*\phi_\beta d\bfy
 =\int_{\calU_0}\widetilde{\phi}_\alpha^*\widetilde{\phi}_\beta d\bfX.
 \label{eq:sm_unitarity}
\eeq
The spatial derivatives satisfy
\beq
\nabla_{\bfy}
=
\bfF(t)^{-T}\nabla_{\bfX},
\label{eq:gradient_transform}
\eeq
and therefore
\beq
\nabla_{\bfy}^{2}
=
\nabla_{\bfX}\cdot \bfF(t)^{-1}\bfF(t)^{-T}\nabla_{\bfX},
\label{eq:laplacian_transform}
\eeq
Differentiating \eqref{eq:wavetrans} at fixed
$\bfX$ gives
\beq
\frac{\partial\widetilde{\phi}_\alpha}{\partial t} = \frac{1}{2}\frac{\dot J}{J}\widetilde{\phi}_\alpha + J^{1/2} \left[\frac{\partial\phi_\alpha}{\partial t} |_{\bm{y}} + (\bfA\bfX)\cdot \nabla_{\bfy}\phi_\alpha \right]_{\bfy=\bfF\bfX}.
\label{eq:chi_time_derivative_derivation}
\eeq
Using \eqref{eq:gradient_transform}, the third term in above equation becomes
\beq (\bfA\bfX)\cdot\nabla_{\bfy} =   (\bfA\bfX)\cdot \bfF^{-T}\nabla_{\bfX} = (\bfF^{-1}\bfA\bfX) \cdot\nabla_{\bfX}.
\label{eq:transport_derivative_derivation}
\eeq
Define the reference frame velocity field
\beq
\bfc(\bfX,t) = \bfF^{-1}\bfA\bfX.
\label{eq:c_definition_derivation}
\eeq
Because $\bfF^{-1}\bfA$ is independent of $\bfX$,
\beq
\frac{\dot J}{J} =
\operatorname{tr}
\left(
\bfF^{-1}\bfA
\right) = \nabla_{\bfX}\cdot\bfc
\label{eq:c_divergence_derivation}
\eeq
Putting these together in  \eqref{eq:chi_time_derivative_derivation}, we get
\beqs
i\frac{\partial \widetilde{\phi}_n}{\partial t}  =  \Bigg[-\frac{1}{2}\nabla_\bfX \cdot \bfG(t)\nabla_\bfX + i\left(\bfc(\bfX,t)\cdot \nabla_\bfX + \frac{1}{2}\nabla_{\bfX}\cdot \bfc(\bfX,t)\right) + \nonumber\\
\widetilde{\varphi}(\bfX,t) + \widetilde{V}_{xc}^{\text{ALDA}}(\widetilde{\rho}_\psi(\bfX,t)) + \widetilde{V}_{\text{NL}}(t) \Bigg]\widetilde{\phi}_n,
\label{eq:lagtddftapp}
\eeqs

\begin{proposition}[Direct discretization of the general Lagrangian TDDFT]
\label{prop:sm_direct_chi_discretization}
Let \(P_\mu^h\) be the centered finite-difference representation of
\(\partial/\partial X_\mu\) after every stencil point outside \(\calU_0\) has
been mapped to the fundamental domain.  Define
\beq
 M_\mu
 =\operatorname{diag}\!\left[
   \bfc(\bfX_g,t)_\mu
  \right],
 \qquad
 A_h=\sum_\mu M_\mu P_\mu^h.
 \label{eq:sm_direct_transport_matrix}
\eeq
Then an adjoint-consistent discrete Hamiltonian for
\eqref{eq:lagtddftapp} is
\beq
 H_\chi^h
 =\frac12\sum_{\mu,\nu}
   (P_\mu^h)^\dagger G_{\mu\nu}(t)P_\nu^h
 +\frac{i}{2}\left(A_h-A_h^\dagger\right)
 +\widetilde{\varphi}^h + \widetilde{V}_{xc}^{\text{ALDA},h} + \widetilde{V}_{\text{NL},h}.
 \label{eq:sm_direct_chi_hamiltonian}
\eeq
For real symmetric \(\bfG\), real \(\widetilde \varphi^h\) and $\widetilde{V}^{\text{ALDA},h}$, and
Hermitian \(\widetilde V_{\chi}^{\mathrm{NL},h}\), the matrix
\(H_\chi^h\) is Hermitian even when \(P_\mu^h\) is not individually
skew-Hermitian after boundary wrapping.
\end{proposition}

\begin{proof}
For a stencil point \(\bfX_{g'}\notin\calU_0\), write
\beq
 \bfX_{g'}=\overline{\bfX}_{g'}+\nu^i\bfe_i,
 \qquad
 \overline{\bfX}_{g'}\in\calU_0,
 \label{eq:sm_chi_ghost_representative}
\eeq
and use the pulled-back image relation
\beq
 \begin{split}
 \widetilde{\phi}_\alpha(\bfX_{g'},t)
 &=J^{1/2}\phi_\alpha\!\left(
   \bfF(t)\overline{\bfX}_{g'}+\bfF(t)\nu^i\bfe_i,t\right)\\
 &=J^{1/2}e^{i\Theta_\nu(\bfF(t)\overline{\bfX}_{g'},t)}
   \phi_\alpha(\bfF(t)\overline{\bfX}_{g'},t)\\
 &=e^{i\eta_\nu(\overline{\bfX}_{g'},t)}
   \widetilde{\phi}_\alpha(\overline{\bfX}_{g'},t),
 \end{split}
 \label{eq:sm_chi_ghost_map}
\eeq
where
\beqs
 \eta_\nu(\bfX,t)
 &=&\Theta_\nu(\bfF(t)\bfX,t) \nonumber\\
 &=&(\nu^i\bfe_i)^T\bfA^T\bfF(t)\bfX
 +\frac12(\nu^i\bfe_i)^T\bfA^T\bfF(t)(\nu^j\bfe_j).
 \label{eq:sm_chi_ghost_phase}
\eeqs
This mapping is used while assembling each \(P_\mu^h\).  The kinetic part of
Eq.~\eqref{eq:sm_direct_chi_hamiltonian} satisfies
\beq
 \begin{split}
 \left[
 \sum_{\mu,\nu}(P_\mu^h)^\dagger
 G_{\mu\nu}P_\nu^h
 \right]^\dagger
 &=\sum_{\mu,\nu}(P_\nu^h)^\dagger
 G_{\mu\nu}P_\mu^h\\
 &=\sum_{\mu,\nu}(P_\mu^h)^\dagger
 G_{\mu\nu}P_\nu^h,
 \end{split}
 \label{eq:sm_direct_kinetic_adjoint}
\eeq
where the second equality follows by interchanging \(\mu\) and \(\nu\) and
using \(G_{\mu\nu}=G_{\nu\mu}\).  Likewise,
\beq
 \left[\frac{i}{2}(A_h-A_h^\dagger)\right]^\dagger
 =\frac{i}{2}(A_h-A_h^\dagger).
 \label{eq:sm_direct_transport_adjoint}
\eeq
The remaining two terms are Hermitian by assumption, proving
\((H_\chi^h)^\dagger=H_\chi^h\).
\end{proof}

The antisymmetrized transport term is not an arbitrary numerical
symmetrization.  If
\(A=\bfc(\bfX,t)\cdot\nabla_{\bfX}\), integration by parts on the reference FD gives
\beq
 A^\dagger=-A-\nabla_{\bfX}\cdot\bfc(\bfX,t)
 =-A-\operatorname{tr}(\bfF^{-1}\bfA).
 \label{eq:sm_continuum_transport_adjoint}
\eeq
Therefore,
\beq
 \frac{i}{2}(A-A^\dagger)
 =i\left(\bfc(\bfX,t)\cdot\nabla_{\bfX}
 +\frac12\operatorname{tr}\bf(\bfF^{-1}\bfA)\right),
 \label{eq:sm_transport_antisymmetrization}
\eeq
which is precisely second term in
Eq.~\eqref{eq:lagtddftapp}.  Discretizing these two terms
separately need not preserve this adjoint identity at wrapped stencil points;
Eq.~\eqref{eq:sm_direct_chi_hamiltonian} enforces their joint structure at
finite grid spacing.

\bibliographystyle{apsrev4-2}
\bibliography{main}

\end{document}

%% file: revtex_preamble.tex
\usepackage{graphics}
\usepackage[caption=false]{subfig}
\usepackage{dcolumn}
\usepackage{bm}
\DeclareMathAlphabet{\mathpzc}{OT1}{pzc}{m}{it}
\usepackage{fullpage}
\usepackage{booktabs}

\usepackage{multirow}
\usepackage{esint}
\usepackage{color}
\usepackage{graphicx}
\usepackage{epstopdf}
\usepackage{verbatim}

\usepackage{fancyvrb}
\usepackage[export]{adjustbox}
\usepackage{svg} 
\usepackage[version=4]{mhchem}
\usepackage{textcomp, gensymb}

\usepackage{textcomp, gensymb}
\usepackage{tikz}
\usepackage{pgfplots}%
\usepackage{pgfplotstable}
\pgfplotsset{compat = newest}
\usetikzlibrary{shapes.geometric, arrows}
\usepackage{placeins}
\usepackage{tabularx}
\usepackage{array}

\usepackage{hyperref}

\usepackage{tikz}
\usetikzlibrary{shapes.geometric, arrows}
\tikzstyle{startstop} = [rectangle, rounded corners, minimum width=3cm, minimum height=1cm,text centered, draw=black, fill={rgb,255:red,197;green,234;blue,250}]
\tikzstyle{io} = [trapezium, trapezium stretches=true, trapezium left angle=70, trapezium right angle=110, minimum width=3cm, minimum height=1cm, text centered, draw=black, fill={rgb,255:red,197;green,234;blue,250}]
\tikzstyle{process} = [rectangle, minimum width=3cm, minimum height=1cm,  draw=black, thick, fill=white]
\tikzstyle{decision} = [diamond, minimum width=3cm, minimum height=1cm, text centered, draw=black, thick,fill=white]
\tikzstyle{arrow} = [thick,->,>=stealth]

\usepackage{mathrsfs}

\usepackage{amssymb}

\makeatletter
\let\cat@comma@active\@empty
\makeatother
\usepackage{breqn}
\usepackage{mathrsfs}

\usepackage[margin=1in]{geometry}
\newcommand{\bfa}{{\bf a}}
\newcommand{\bfb}{{\bf b}}
\newcommand{\bfc}{{\bf c}}

\newcommand{\bfe}{{\bf e}}
\newcommand{\bff}{{\bf f}}

\newcommand{\bfh}{{\bf h}}

\newcommand{\bfv}{{\bf v}}

\newcommand{\bfx}{{\bf x}}
\newcommand{\bfy}{{\bf y}}

\newcommand{\bfA}{{\bf A}}
\newcommand{\bfB}{{\bf B}}

\newcommand{\bfF}{{\bf F}}
\newcommand{\bfG}{{\bf G}}

\newcommand{\bfI}{{\bf I}}

\newcommand{\bfQ}{{\bf Q}}

\newcommand{\bfT}{{\bf T}}

\newcommand{\bfX}{{\bf X}}

\newcommand{\beq}{\begin{equation}}
\newcommand{\eeq}{\end{equation}}
\newcommand{\beqs}{\begin{eqnarray}}
\newcommand{\eeqs}{\end{eqnarray}}
\newcommand{\beql}{\begin{equation} \label}

\newcommand{\calA}{{\cal A}}

\newcommand{\calG}{{\cal G}}

\newcommand{\calU}{{\cal U}}

\newcommand{\Z}{\mathbb{Z}}
\newcommand{\R}{\mathbb{R}}
\newcommand{\C}{\mathbb{C}}

\let\oldFootnote\footnote
\newcommand\nextToken\relax

\renewcommand\footnote[1]{%
    \oldFootnote{#1}\futurelet\nextToken\isFootnote}

\newcommand\isFootnote{%
    \ifx\footnote\nextToken\textsuperscript{,}\fi}

\usepackage{amsthm}
\newtheorem{definition}{Definition}
\newtheorem{proposition}{Proposition}
\newtheorem{lemma}{Lemma}
\newtheorem{remark}{Remark}
\newtheorem{theorem}{Theorem}